\pdfoutput=1
\documentclass[preprint,showpacs,showkeys,preprintnumbers,aps,amssymb]{revtex4}
\usepackage{graphicx}
\usepackage{amsmath}
\usepackage{bm}
\usepackage[colorlinks=true, pdfstartview=FitV, linkcolor=red, citecolor=blue, urlcolor=blue, pdftitle={Small shear viscosity in the semi quark gluon plasma},pdfauthor={Yoshimasa Hidaka and Robert D. Pisarski},pdfsubject={}, pdfkeywords={Heavy ion collisions, shear viscosity, quark gluon plasma}]{hyperref}
\newcommand{\Nc}{N_c}
\newcommand{\Nf}{N_f}
\newcommand{\Tc}{T_c}
\newcommand{\slat}{s_\text{lat}}
\newcommand{\mTsemi}{T^-_{\rm semi}}
\newcommand{\pTsemi}{T^+_{\rm semi}}
\newcommand{\SU}{SU}  
\newcommand{\tr}{\mathrm{tr}}
\newcommand{\psibar}{\overline{\psi}}
\newcommand{\Acl}{A_{0}^{\text{cl}} }
\newcommand{\bx}{\bm{x}}

\newcommand{\bk}{\bm{k}}
\newcommand{\bp}{\bm{p}}

\newcommand{\ploop}{L}
\newcommand{\ratioR}{\mathcal{R}(\ell)}
\newcommand{\collisionOperator}{\mathcal{L}}
\newcommand{\Slash}[1]{\ooalign{\hfil/\hfil\crcr$#1$}}
\newcommand{\arXiv}[2]{\href{http://arxiv.org/abs/#1}{#2}}
\newcommand{\doi}[2]{#2}

\newcommand{\Lag}{\collisionOperator}

\def\anp#1#2#3{Annals Phys. {\bf #1}, #2 (#3)}
\def\arnps#1#2#3{Ann.\ Rev.\ Nucl.\ Part.\ Sci.\  {\bf #1}, #2 (#3)}
\def\atmp#1#2#3{Adv. Theor. Math. Phys. {\bf #1}, #2 (#3)}
\def\cmp#1#2#3{Comm. Math. Phys. {\bf #1}, #2 (#3)}

\def\cqg#1#2#3{Class.\ Quant.\ Grav.\ {\bf #1}, #2 (#3)}
\def\epja#1#2#3{Eur.\ Phys.\ Jour.\ A{\bf #1}, #2 (#3)}
\def\epjc#1#2#3{Eur.\ Phys.\ Jour.\ C{\bf #1}, #2 (#3)}
\def\epl#1#2#3{Eur.\ Phys.\ Lett {\bf #1}, #2 (#3)}
\def\ibid#1#2#3{{\it ibid.} {\bf #1}, #2 (#3)}

\def\ijme#1#2#3{Int. Jour. Mod. Phys. E {\bf #1}, #2 (#3)}
\def\jhep#1#2#3{Jour. High Energy Phys. {\bf #1}, #2 (#3)}

\def\jph#1#2#3{Jour. of Phys. {\bf #1}, #2 (#3)}
\def\npa#1#2#3{Nucl. Phys. A {\bf #1}, #2 (#3)}
\def\npb#1#2#3{Nucl. Phys. B {\bf #1}, #2 (#3)}

\def\plb#1#2#3{Phys. Lett. B {\bf #1}, #2 (#3)}
\def\prc#1#2#3{Phys. Rev. C {\bf #1}, #2 (#3)}
\def\prd#1#2#3{Phys. Rev. D {\bf #1}, #2 (#3)}
\def\prl#1#2#3{Phys. Rev. Lett. {\bf #1}, #2 (#3)}
\def\phr#1#2#3{Phys. Rep. {\bf #1}, #2 (#3)}

\def\ptp#1#2#3{Prog. Theor. Phys. {\bf #1}, #2 (#3)}

\def\rpp#1#2#3{Rept. Prog. Phys. {\bf #1}, #2 (#3)}
\def\rmp#1#2#3{Rev. Mod. Phys. {\bf #1}, #2 (#3)}

\def\zpc#1#2#3{Z. Phys. C {\bf #1}, #2 (#3)}

\begin{document}
\preprint{KUNS-2242}
\preprint{BNL-90640-2009-JA}
\title{Small shear viscosity in the semi quark gluon plasma}
\author{Yoshimasa Hidaka$^{a}$
and Robert D. Pisarski$^{b}$}
\affiliation{
$^a$Department of Physics, Kyoto University, Sakyo-ku, Kyoto 606-8502, Japan\\
$^b$Department of Physics, Brookhaven National Laboratory, Upton, NY 11973, USA\\
}
\date{\today}
\begin{abstract}
At nonzero temperature in QCD,
about the deconfining phase transition 
there is a ``semi'' quark gluon plasma (semi-QGP), where
the expectation value of the (renormalized) Polyakov loop is less than one.
This can be modeled by a semiclassical
expansion about
a constant field for the vector potential, $A_0$, which is 
diagonal in color.  We compute the shear viscosity in
the semi-QGP by using the Boltzmann equation in the presence of this
background field.  
To leading, logarithmic order in weak coupling, the dominant
diagrams are given by the usual scattering processes of
$2 \rightarrow 2$ particles.  For simplicity
we also assume that both the number of colors and flavors are large.
Near the critical temperature $\Tc$,
where the expectation value of the Polyakov loop is small,
the overall density of colored fields decreases according to their color
representation, with
the density of quarks vanishes linearly with the loop, and that of gluons,
quadratically.
This decrease in the overall density dominates
changes in the transport cross section.  As a result,
relative to that in the perturbative QGP, near $\Tc$
the shear viscosity in the semi-QGP is suppressed by two powers of
the Polyakov loop.
In a semiclassical expansion, the suppression of colored fields depends
only upon which color representation they lie in, and not upon their mass.
That light and heavy quarks are suppressed in a common manner may help
to explain the behavior of charm quarks at RHIC.
\end{abstract}

\maketitle
\section{Introduction}

The collisions of heavy ions at ultrarelativistic energies have exhibited
a multitude of 
surprising results \cite{whitepaper,strong,strong_rdp,hydro}.  Experiments 
at the Relativistic Heavy Ion Collider (RHIC) at Brookhaven have shown that in
peripheral collisions, the anisotropy with respect to the reaction
plane, or 
elliptic flow, can described by nearly ideal hydrodynamics, with
a small value for the shear viscosity.
In kinetic theory, 
the shear viscosity is inversely proportional to the scattering
amplitude \cite{PhysicalKinetics},
so a small shear viscosity could be due to
strong coupling.  This suggests that is fruitful to consider
the analogy to  ${\cal N}=4$ supersymmetric gauge theories,
where one can compute at infinite coupling for an
infinite number of colors.  For ${\cal N}=4$ supersymmetry,
the ratio of the shear viscosity to the entropy density is
$1/4 \pi$, which is conjectured to be a universal lower bound 
\cite{susy1,susy2}.  
The shear viscosity has also been computed from
numerical simulations on the lattice~\cite{lattice}, 
in the hadronic phase~\cite{hadronic},
and in the perturbative quark gluon 
plasma (QGP)~\cite{transport1,baym,transport2,Jeon,Gagnon,diagram,amy,Asakawa,transport3,lebellac}.

In this paper we adopt an alternate approach
from ${\cal N}=4$ supersymmetry.
Instead of trying to work down from infinite
coupling, we work up from small coupling.  While perturbation theory
at nonzero temperature is badly behaved, resummed perturbation theory
works down to much lower temperatures, to a few times $\Tc$,
where $\Tc$ is the critical temperature for deconfinement
\cite{pert_calcs,braaten,pert_review,resum,coupling}.
This suggests that nonperturbative effects dominate the 
region near $\Tc$,
which we have termed the semi-QGP 
\cite{loopRPa,loopRPb,loopOther,lattice_effective,pnjl}.
In this paper, the fourth in a series \cite{yk_rdp1,yk_rdp2,yk_rdp3},
we compute the shear viscosity in a simple approximation 
for the semi-QGP.

The basis of this approach are measurements of the order parameters
for deconfinement.  Consider a straight Wilson line
in the direction of imaginary time,
\begin{equation} 
\ploop = P \exp \left( i\int_0^{1/T} d\tau A_\tau \right) \,,
\label{eq:PolyakovLoop}
\end{equation}
where $P$ denotes path ordering, $T$ is the temperature,
$\tau$ the imaginary time, and $A_\tau$ is the timelike 
component of the gauge field, in the fundamental representation.
The Wilson line is a matrix in color space, and so is not gauge invariant,
but its eigenvalues are.  The simplest measure of the eigenvalues of the
Wilson line is its trace, which is the Polyakov loop,
\begin{equation}
\ell = \frac{1}{\Nc}\; {\rm tr} \; \ploop \; .
\label{loop_definition}
\end{equation}
This Polyakov loop is directly related 
to the propagator of an infinitely heavy quark, which for
Eq.~(\ref{loop_definition}) is
in the fundamental representation.
Heuristically, one can
view the Polyakov loop as measuring the excess free energy $f_q$
which arises from adding a colored, heavy 
quark to a thermal bath,
$\langle \ell \rangle \sim \exp(-f_q/T)$.  

This Polyakov loop is a bare quantity, but a renormalized quantity can
be extracted from simulations on the lattice 
\cite{yk_rdp3,renloop_pert,ren_lat,cheng,detar}.  These indicate that the 
expectation value of the Polyakov loop in Eq.~(\ref{loop_definition})
is approximately constant and near one for temperatures greater than
$\sim 3 \, \Tc$.  
Below $\sim 3 \, \Tc$, however, the Polyakov loop is significantly
less than one.  
In a pure gauge theory, the Polyakov loop is a strict order parameter for
deconfinement, so below $\Tc$ the excess free energy
$f_q$ is infinite, and the expectation value of the
Polyakov loop vanishes identically.
With dynamical quarks, the Polyakov loop is only an approximate
order parameter: because of screening by quark antiquark pairs,
below $\Tc$ the excess free energy
$f_q$ is finite, and $\langle \ell \rangle$ is nonzero.
In a gauge theory with three colors and three light flavors,
as in QCD, numerical simulations find that the Polyakov loop
is small at $\Tc$, $\langle \ell \rangle \approx 0.2$,
and essentially vanishes below $\approx 0.8 \, \Tc$;
see, {\it e.g.}, 
Fig.~(13) of Bazavov {\it et al.}, \cite{ren_lat,cheng}.  We 
dub the region in which
the Polyakov loop deviates from one as the ``semi''-QGP 
\cite{yk_rdp1,yk_rdp2,yk_rdp3}.

Similar considerations have also motivated what is known at the 
Polyakov--Nambu--Jona-Lasino model \cite{pnjl}.  In these models, the only
dynamical variable related to deconfinement is the simplest Polyakov loop,
that in the fundamental representation, Eq.~(\ref{loop_definition}).
For an $\SU(\Nc)$ gauge theory, however, the thermal Wilson line has
$\Nc - 1$ independent eigenvalues.  
These are characterized
by traces of higher powers of $\ploop$, ${\rm tr} \, \ploop^2$
through ${\rm tr} \, \ploop^{\Nc - 1}$.
These higher moments of the thermal Wilson line are not directly
accessible on the lattice, since loops in different representations
have distinct renormalization constants.
These loops have a simple physical interpretation, as the propagator of
a heavy quark in that representation.  

In the end, we parametrize our results in terms of the Polyakov loop
in the fundamental representation.  Even so, in intermediate steps,
we could not represent the computations by a simplified model in which
only the simplest Polyakov loop enters: the matrix structure is essential.
We discuss this further in the Conclusions, Sec.~\ref{bleaching}.

Our approach to the semi-QGP is the following:  To the classical
Lagrangian, we add terms which drive confinement, such as \cite{loopRPb}
\begin{equation}
{\cal L}_\text{eff} = \; b_\text{fuzzy} \; T^2 \; 
\Tc^2 \; | \tr \; \ploop |^2 \,;
\label{fuzzy}
\end{equation}
there are many other possible terms \cite{loopRPa,loopRPb,loopOther}.  
To date, a simple
form for the effective Lagrangian for the semi-QGP has not been obtained;
this would allow one to compute both the pressure and the renormalized 
loop from the same effective Lagrangian \cite{unpub}.  Clearly
analysis from numerical simulations, especially in effective models,
is essential to gaining this understanding \cite{lattice_effective}.

We shall see that knowing the full form for the effective Lagrangian
is irrelevant to the question we address in this paper: near $\Tc$,
where the expectation value of the Polyakov loop is small, 
is the shear viscosity suppressed, or enhanced?  Naive expectation, based
upon kinetic theory, indicates that as the cross section decreases,
the shear viscosity increases.  
In contrast, we find that the shear viscosity in the semi-QGP
is smaller than in the perturbative QGP.

We note that 
our approach to the semi-QGP breaks strong resemblance to ``double-trace''
deformations of the vacuum theory \cite{unsal}.  In these models,
a term such as Eq.~(\ref{fuzzy}) is added to drive the theory into a
confining phase.  

This paper is organized as follows:
In Sec.~\ref{sec:effectiveTheory}, we consider a semiclassical
expansion of the semi-QGP.  For technical reasons we
assume that both $\Nc$ and $\Nf$ are large.
In Sec.~\ref{transPortCoefficient}, we discuss how
to compute transport coefficients in the 
semi-QGP using kinetic theory.
This is a standard approach, 
except that kinetic theory is in the presence
of a background field, $A_\tau$ in Eq.~(\ref{loop_definition}). 
This background field is important near the phase transition,
and suppresses the shear viscosity.
Section~\ref{sec:NumericalResults} gives numerical results.
In Sec.~\ref{bleaching} we discuss possible phenomenological
implications of our results.  We find that in
a semiclassical expansion of the semi-QGP, 
near $\Tc$ all colored particles
are suppressed in a universal manner.  This is natural, as a type
of ``bleaching'' of color as $T \rightarrow T_c^+$.  
In particular, this color bleaching is independent of the mass
of the field.  We suggest that this might help to explain the otherwise
puzzling results on the behavior of charm quarks at RHIC.

\section{Effective theory in the Semi-QGP}
\label{sec:effectiveTheory}
\begin{figure}
\resizebox{0.45\textwidth}{!}{%
  \includegraphics{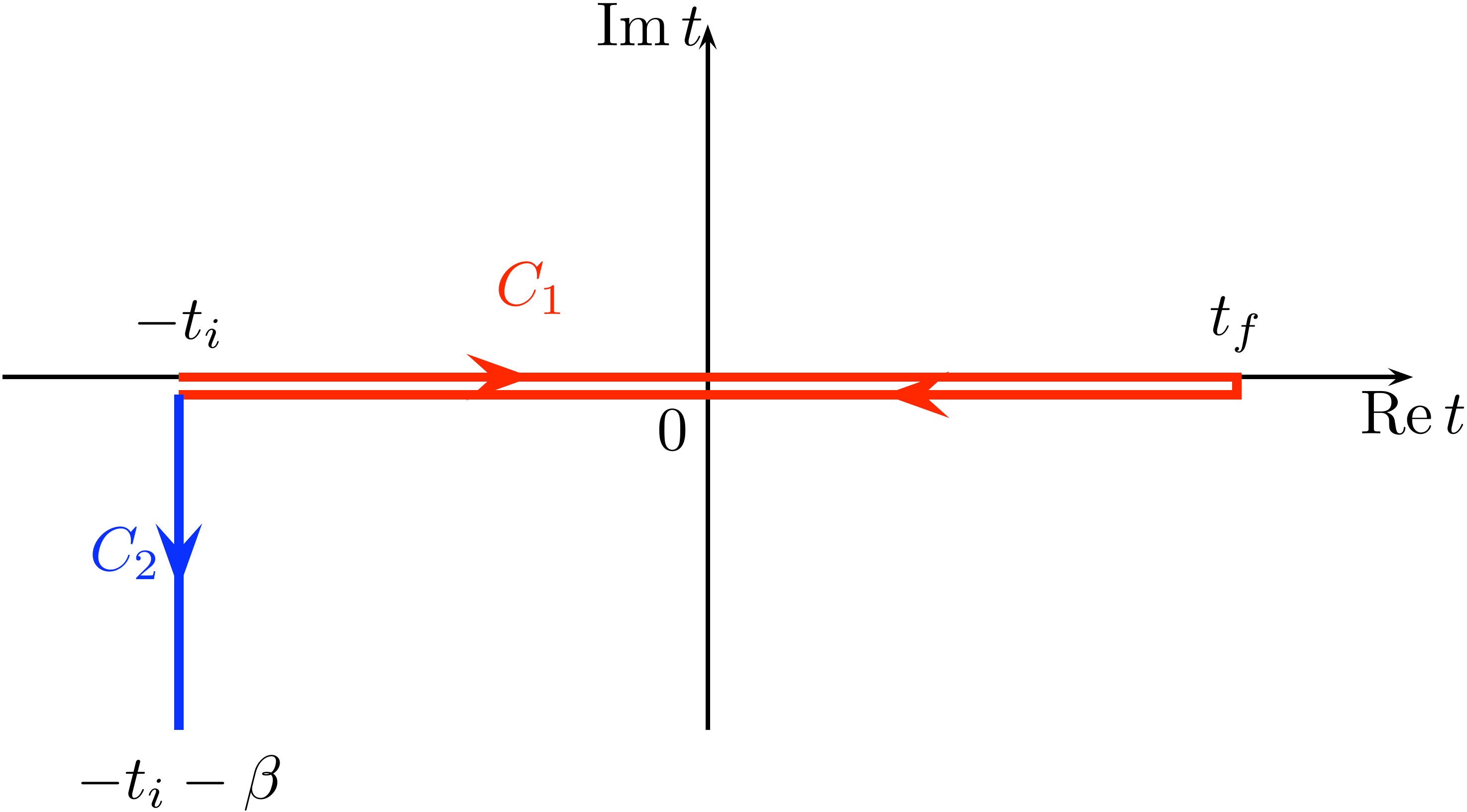}
}
\caption{Complex-time path.}
\label{timepath}
\end{figure}

A semi-QGP is characterized by the Polyakov loop, 
Eq.~(\ref{loop_definition}).  
Static quantities, such as the pressure,
are determined by an effective Lagrangian of Polyakov 
loops~\cite{loopRPa,loopRPb,loopOther,lattice_effective,pnjl}.
What we require is an effective theory in real time,
which can be used to compute transport coefficients.
Start with the partition function for QCD,
\begin{equation}
Z(T) 
 =\int\mathcal{D}A_\mu \, \mathcal{D}\psi \, \mathcal{D}\psibar 
\; \exp[i\int_C d^4x \, \Lag] \,,
\end{equation}
where $\Lag$ is the Lagrangian,
\begin{equation}
\Lag=-\frac{1}{2}\, \tr \, F_{\mu\nu}^2 + \psibar (i\Slash{D}-m)\psi  \,,
\end{equation}
with $F_{\mu\nu}=\partial_\mu A_\nu-\partial_\nu A_\mu -ig[A_\mu, A_\nu]$;
$\psi$ and $\psibar$ is the quark field, 
$\Slash{D}=\gamma^\mu(\partial_\mu-igA_\mu)$, and $m$ is the quark mass.
$C$ denotes a path in complex time, Fig.~(\ref{timepath}).
We work in Minkowski 
spacetime with a metric, $g_{\mu\nu}=g^{\mu\nu}=\text{diag}(1,-1,-1,-1)$.
The timelike gauge field on the imaginary time in 
Eq.~(\ref{eq:PolyakovLoop}), $A_{\tau}$, corresponds to
$i A_{0}$ in this notation.
We defer a discussion of a fully self-consistent approach, where terms
such as Eq.~(\ref{fuzzy}) are added, to later analysis.  
Including such a term would not alter our results at leading order in
$g$, because to leading order the shear viscosity is
dominated by the scattering of hard particles, with momenta $\sim T$,
while terms at Eq.~(\ref{fuzzy}) affect fields at soft momenta.

For the generators of $SU(\Nc)$, we use what may be called
't Hooft's double line basis~\cite{hooft,yk_rdp2}:
\begin{equation}
[t^{ab}]_{cd} =\frac{1}{\sqrt{2}}\Bigl(\delta^a_c \, \delta^b_d  
-\frac{1}{\Nc}\, \delta^{ab}\, \delta_{cd} \Bigr).
\label{basis1}
\end{equation}
Somewhat unconventionally, we denote indices in the fundamental representation
as $a, b,c,d,\ldots = 1\ldots\Nc$.  In the double line basis, generators
are denoted by a pair of indices in the fundamental representation,
$ab$, so that in all there are $\Nc^2$ generators.  This is one more
generator than in an orthonormal basis.  As an overcomplete basis, 
the trace of two generators is not a delta function, but
a projection operator:
\begin{equation}
\mathrm{tr}\, t^{ab}\, t^{cd}=\frac{1}{2} \, P^{ab,cd}
\;\; , \;\; P^{ab,cd} =
\delta^{ad}\, \delta^{bc}-\frac{1}{\Nc}\, \delta^{ab}\, \delta^{cd} \,.
\label{basis2}
\end{equation}
The commutator of two generators is
\begin{equation}
[t^{ab},t^{cd}] = i\sum_{e,f=1}^{\Nc}\,f^{(ab,cd,ef)}\, t^{fe} \,,
\end{equation}
where $f^{(ab,cd,ef)}$ is the structure constant,
\begin{equation}
f^{(ab,cd,ef)}=\frac{i }{\sqrt{2}}\,
\left(\delta^{ad}\, \delta^{cf}\, \delta^{eb}-
\, \delta^{af}\, \delta^{cb}\, \delta^{ed} \right) \,.
\end{equation}
The double line basis is obviously convenient at large $\Nc$, as then the
terms $\sim 1/\Nc$ in Eqs.~(\ref{basis1}) and (\ref{basis2}) can be
dropped.  It is also of use at finite $\Nc$, however, since then
each line in a generator represents a flow of color charge,
and charge conservation is obvious; this is especially true
in the presence of a background field.~\cite{yk_rdp2}.

We denote four momenta as $P^\mu=(p^0,\bp)$,
where $p_0$ is the timelike component of the momenta, and $\bp$
the spatial momentum.  In thermal equilibrium 
at a temperature $T$, 
$p_0 = i \omega_n$, where $\omega_n$ is a Matsubara frequency: 
$\omega_n= 2 n \pi T$ for bosons, and
$= (2n+1) \pi T$ for fermions, where $n$ is an integer.
In imaginary time, a nonzero value of the Polyakov loop is modeled
by taking a constant field for $A_0$, 
\begin{equation}
[\Acl]_{ab} =
i \delta_{a b} \; \frac{ Q^{a}}{g} \; .
\label{ansatz}
\end{equation}
We expand about the classical field, $\Acl$, 
in fluctuations, $B_\mu$:
\begin{equation}
A_{\mu} = \delta_{\mu 0} \; \Acl + B_{\mu} \, .
\end{equation}
The covariant derivatives in this particular background field are
especially simple:
\begin{align}
[i(\partial_0 -ig\Acl) \psi]_{a}&\to (p_0 +iQ^a )\psi_a \; ; \notag\\
[i(\partial_0 -ig[\Acl,.]) B_\mu]_{ab}&\to (p_0 +iQ^{ab} )[B_\mu]_{ab} 
\;\; ; \;\; Q^{ab}\!=\!Q^a\!-\!Q^b \; .
\end{align}
Note that the adjoint covariant derivative involves
the quantity $Q^{a b}$, which is the difference of two
color charges.
The corresponding statistical distribution functions are
\begin{align}
n_a(E)&=\frac{1}{{e}^{|E - i Q^a|_R/T} + 1} \qquad\text{for quarks}, \notag\\
n_{ab}(E)&=\frac{1}{{e}^{|E - i Q^{ab}|_R/T} - 1} \qquad\text{for gluons},
\label{stat_dist_fncs}
\end{align}
where $|\cdots|_{R}$ is defined as
\begin{equation}
|z|_R=
\left\{    \begin{array}{ll}
+z   \quad& \mathrm{Re}\; z>0 \\
 -z   \quad& \mathrm{Re}\; z <0
   \end{array}
   \right. \,.
\end{equation} 
From the statistical distribution functions we see that the background
field acts like an imaginary chemical potential for color.  
As such, in a given field $Q^a$ these distribution functions
are complex, and so unphysical.  
The only physically meaningful quantities are integrals over 
distributions of $Q$, and these give results which are sensible.

To see how this comes about, consider summing a quark
propagator over its color indices.  This enters, for example, in the
computation of the pressure at leading order.  The sum is
\begin{equation}
\frac{1}{\Nc}\sum_{a}\frac{1}{{e}^{(E - i Q^a)/T} + 1}
= \sum_{n=1}^{\infty} (-)^{n+1} \;
{e}^{- n E /T}\; \frac{1}{\Nc}\; \tr\; \ploop^n  \,.
\label{expansion}
\end{equation} 
We can always assume that the vacuum expectation value of $\tr\, \ploop$
is real.  In the pure glue theory, this can be enforced by a global
$Z(\Nc)$ rotation; with dynamical quarks, this is automatic.  In
Eq.~(\ref{expansion}) there is an infinite series
in powers of $\tr\, \ploop^n$, the expectation values of which are real.
The computation of the shear viscosity in this paper provides another
example where integrals over $Q$'s give physically sensible results.

We wish to compute correlation functions near thermal equilibrium,
and so need to continue from imaginary to real time.  We adopt the
usual path, illustrated in Fig.~(\ref{timepath}), along
$C=C_1\cup C_2$.  This includes integration in real time, $\mathrm{Re}\, t$,
along $C_1$ 
from an initial time $t_i$, to a final time, $t_f$, and back (in practice,
both times are assumed to be infinite).  
Then one integrates in imaginary time, along $C_2$ 
from $\mathrm{Im}\, t: 0 \rightarrow -1/T = -\beta$.

It is natural to take the background field only for the part of the path
in imaginary time.  Consider an ordinary chemical potential,
introduced as a Lagrange multiplier for the number
operator.  The chemical potential does not affect the Hamiltonian, nor the
evolution in real time: it enters only to change the statistical distribution
functions in thermal equilibrium.  Thus we do the same for an imaginary
chemical potential for color: along
Fig.~(\ref{timepath}) we take the background field to be nonzero
along $C_2$, and to vanish along $C_1$.

While we will not use the real time formalism, it helps to understand
the choice of background field in time.
The propagator is a two by two matrix,
\begin{equation}
[G_{\mu\nu}(K)]_{ab,cd}=-g_{\mu\nu}\;
P_{ab,cd}\; D_{ab}(K) \,,
\end{equation}
where
\begin{equation}
D_{ab}(K) = 
\begin{pmatrix}
D_{ab}^{11}(K) & D_{ab}^{12}(K) \\
D_{ab}^{21}(K) & D_{ab}^{22}(K) 
\end{pmatrix}  
\end{equation}
with
\begin{align}
    D_{ab}^{11}(K)   =& \frac{i}{K^2+i\epsilon}+n_{ab}(k_0)\;
2\pi\; \delta(K^2)  \,, \notag\\
    D_{ab}^{12} (K)  =& (n_{ab}(k_0) +\theta(-k_0))\;
2\pi\; \delta(K^2) \,, \notag\\
    D_{ab}^{21} (K)  =&(n_{ab}(k_0) +\theta(k_0))\;
2\pi\; \delta(K^2)  \,, \notag\\
    D_{ab}^{22} (K)  =&   \frac{-i}{K^2-i\epsilon}+n_{ab}(k_0)\;
2\pi\; \delta(K^2) \,,
\label{real_time_dist_fncs}
\end{align}
in Feynman gauge~\cite{Furuuchi:2005zp},
where $\theta(k^{0})$ is the Heaviside step function.
This form shows that the zero temperature parts of
the propagator, $1/(K^2 \pm i \epsilon)$,
are independent of the $Q$'s.  The color charges only enter
through the statistical distribution functions, $n_{ab}(k_0)$.

Instead of using the real time formalism, we compute in imaginary time,
and then analytically continue.  The 
color dependent frequency $\omega_n$ becomes
\begin{equation}
i\omega_n+iQ^{ab} \to \omega \pm i\epsilon \, ,
\end{equation}
where $\pm$ corresponds to retarded and advanced propagation.  

After computing diagrams, we need to sum over distributions of the $Q$'s.
To avoid having to deal with trace terms at finite $\Nc$, we work
in the limit of large $\Nc$.  The pressure is of order $\sim \Nc^0$
in the hadronic phase, and $\sim \Nc^2$ in the deconfined phase.
Thus a mean field approximation,
as in Eq.~(\ref{ansatz}), should be a reasonable approximation
in the deconfined phase at large $\Nc$.
At large $\Nc$, color sums can be replaced by integrals:
\begin{equation}
\frac{1}{\Nc}\sum_{a}\to\int da = \int_{-\pi}^{\pi} d\varphi \; \rho(\varphi) \,,
\end{equation}
where we introduce the eigenvalue density
$\rho(\varphi)=da/d\varphi$, with $\varphi=Q/T$.
Moments of the Polyakov loops are 
Fourier transforms of the eigenvalue density,
\begin{equation}
\ell_n=\frac{1}{\Nc}\; \tr \ploop^n =
\int_{-\pi}^\pi d\varphi \; \rho(\varphi) \; e^{i\varphi n} \, .
\end{equation}
In particular, we write the 
the Polyakov loop in the fundamental representation, $\ell_{1}$,
simply as $\ell$.
In the confined phase, all moments of the Polyakov loop vanish,
$\ell_{n}=0$, and the eigenvalue density is constant.
In the perturbative QGP, the moments of all Polyakov loops are
close to one, $\ell_{n}=1$, and the eigenvalue density
is close to a delta function.

Since we do not now have a complete effective theory of the semi-QGP,
we do not know the exact eigenvalue density as a function of temperature.
We do know the expectation value of the simplest Polyakov loop,
$\ell$, as a function of temperature, and so we
consider two representative forms for the eigenvalue density.
One ansatz is to take a step function,
\begin{equation}
\rho_\text{step}(\varphi,\lambda) = \frac{1}{2\lambda}\theta(\lambda-|\varphi|) \,,
\label{eq:distributionStep}
\end{equation}
where $\lambda$ is a parameter characterizing the eigenvalue density. 
With this ansatz, the expectation value of the Polyakov loops are
\begin{equation}
\ell_n \equiv \frac{1}{\Nc} \; \tr \; \ploop^n
=\frac{\sin n\lambda}{n \lambda} \,.
\end{equation} 

The other ansatz we take from a soluble model in two dimensions,
that of Gross and Witten \cite{gross}:
\begin{equation}
\rho_\text{GW}(\varphi,\lambda) =\left\{    \begin{array}{ll}
 \frac{1}{2\pi}\left( 1+\lambda\cos\varphi \right)&\lambda\leq 1\\
     \frac{\sqrt{\lambda}}{\pi}\cos\frac{\varphi}{2} \left(1-\lambda\sin^2\frac{\varphi}{2}\right)^{1/2}   
     \quad& 
          \lambda>1
   \end{array}
   \right. \,,
   \label{eq:distributionGW}
\end{equation}
where $ |\varphi|<2\sin^{-1}\left({1}/{\lambda}\right)^{1/2}$ is 
satisfied for $\lambda>1$.
For $\lambda\leq1$, $\ell_{\pm 1}=1/(2\lambda)$, while
all others vanish.
With this ansatz, there is no simple expression
for $\ell_n$ when $\lambda>1$.

Rather unexpectedly, we shall see that our results for the shear viscosity
are almost independent of the choice of these eigenvalue densities.
We do not know if this is a property special to the shear viscosity, or
if it is generic.

\section{Transport coefficients}
\label{transPortCoefficient}
In this section we consider transport coefficients in a semi-QGP.
Hydrodynamics is described by 
evolution in time of locally conserved currents, 
such as the energy-momentum tensor, which 
satisfy $\partial_\mu T^{\mu\nu}=0$. 
For simplicity, we assume that the only
conserved charges are those for energy and momentum.
In hydrodynamics the energy-momentum tensor $T^{\mu\nu}$ can be expanded 
as \cite{FluidDynamics}
\begin{equation}
T^{\mu\nu}= \mathcal{E} u^\mu u^\nu + 
\mathcal{P}\varDelta^{\mu\nu}+\varPi^{\mu\nu} \,,
\label{eq:Tmunu}
\end{equation}
where $u^\mu$ is the local velocity, $u_{\mu}u^{\mu}=1$;
$\mathcal{E}=T^{\mu\nu}u_\mu u_\nu$ is the energy density and
$\mathcal{P}$ is the local pressure, both of which are Lorentz scalars;
lastly, $\varDelta^{\mu\nu}$ is a spatial projection operator,
\begin{equation}
\varDelta^{\mu\nu}=u^\mu u^\nu-g^{\mu\nu} \,.
\label{eq:projection}
\end{equation}
For the metric we take $g_{\mu\nu}=g^{\mu\nu}=\text{diag}(1,-1,-1,-1)$.

$\varPi^{\mu\nu}$ measures the deviation from thermal equilibrium.
Near thermal equilibrium it
can be expanded in a gradient expansion,
\begin{equation}
\varPi^{\mu\nu}=\eta\; \sigma^{\mu\nu}
 +\zeta \; \varDelta^{\mu\nu} \varDelta^{\alpha\beta}
\; \partial_\alpha u_\beta
+\cdots \,,
\label{Tij1}
\end{equation}
where
\begin{equation}
\sigma^{\mu\nu}=\varDelta^{\mu\rho}\varDelta^{\nu\sigma}
\Bigl(\partial_\sigma u_\rho+\partial_\rho u_\sigma-\frac{2}{3}\;
g_{\sigma\rho}\; \varDelta^{\alpha\beta}\partial_\alpha u_\beta\Bigr)  \,.
\end{equation}
The first order transport coefficients $\zeta$ and $\eta$ are,
respectively, the bulk and shear viscosities.
The transport coefficients reflect microscopic properties of the theory,
and are an input into a hydrodynamical analysis.

These transport coefficients can be computed from
the Kubo formula, which relates them to 
the correlation functions \cite{KuboFormula}:
\begin{align}
\eta = &\frac{1}{20}\lim_{\omega\to0}\frac{1}{\omega}
\int d^4x e^{i\omega t}\theta(t) \langle [\pi_{ij}(x),\pi^{ij}(0)]\rangle \,,  \\
\zeta = &\frac{1}{2}\lim_{\omega\to0}
\frac{1}{\omega}\int d^4x e^{i\omega t} \theta(t)
\langle [\mathcal{P}(x),\mathcal{P}(0)]\rangle \, ,
\end{align}
where $\mathcal{P}(x)=-{T^{i}}_{i}(x)/3$ and 
$\pi_{ij}(x)=T_{ij}(x)+g_{ij}\mathcal{P}(x)$.
These correlation functions are not simple to compute in perturbation
theory, though.  
They are sensitive to pinch singularities, and it is necessary
to resum an infinite series of diagrams.

Another way of computing these transport coefficients
is to use kinetic theory and the
Boltzmann equation.  To leading order, this is equivalent to a resummation
of the relevant diagrams in the Kubo formula ~\cite{Jeon,Gagnon}.
We will use the Boltzmann equation 
to compute the shear viscosity in a semi-QGP in ~Sec.~\ref{sec:derivation}.
Our analysis is a straightforward extension of the computation of
Arnold, Moore, and Yaffe \cite{amy}, of the shear
viscosity in the perturbative QGP, to the semi-QGP.

\subsection{Shear viscosity in the semi-QGP} \label{sec:shearViscosityKineticTheory}
Before going to the technical details of the computation,
we give a qualitative discussion of how the shear viscosity
arises in kinetic theory in the QGP, 
and how it differs in the semi-QGP.

For the transport theory of a relativistic gas ~\cite{deGroot},
the shear viscosity $\eta$ is given as 
\begin{equation}
\eta\approx \frac{4}{15}\; n\,  \bar{p} \, \lambda_\text{mfp} \,,
\label{eq:viscosityClassical}
\end{equation}
where $n$ is the number density, $\bar{p}$ is the mean
momentum, and $\lambda_\text{mfp}$ is the 
mean free path. 
For light particles, where the mass is much less than the temperature,
the mean momentum $\bar{p} \sim T$, and the number density
is $n \sim T^3$.
The mean free path is 
$\lambda_\text{mfp}\sim 1/(n\sigma)$, 
where $\sigma$ is the transport cross section, so 
that the shear viscosity is $\eta\sim T/\sigma$.

In QCD, the transport cross section is 
$\sigma\sim g^{4}\ln(T/m_\text{Debye})/T^{2}$, where 
$m_\text{Debye} \sim \sqrt{g^{2}n/T}\sim gT$
is the Debye mass. There is a Coulomb logarithm, $\ln (T/m_\text{Debye})$, 
due an infrared singularity 
from forward scattering~\cite{ichimaru}, so in all,
\begin{equation}
\eta \sim \frac{T^{3}}{g^{4}\, \ln(1/g)} \; .
\label{pert_shear_viscosity}
\end{equation}
The Boltzmann equation gives the same 
powers of $g$ and $\ln\,(1/g)$, but a strictly perturbative analysis
gives the wrong overall coefficient.  To get the correct result 
perturbatively requires the resummation of an infinite set of diagrams
~\cite{baym,amy}.

In weak coupling, $\sigma$ is small and the shear viscosity is large.
This sounds counter intuitive, but is not: 
when perturbed, it takes a long time
for a weakly interacting gas to move back towards thermal
equiblibrium.

Conversely, the most natural way to obtain a small shear viscosity
is if the cross section, and so the coupling constant, are large.
The best illustration of this is a
${\cal N}=4$ supersymmetric gauge theory,
where in the limit of infinite
coupling and an infinite number of colors, the ratio
of the shear viscosity to the entropy density, $s$,
is $\eta/s = 1/4 \pi$ \cite{susy1}.  This is conjectured to
be a universal lower bound on $\eta/s$.

The analysis in a semi-QGP is distinct from ordinary kinetic theory.
It is characterized by the partial ionization of color
\cite{yk_rdp1,yk_rdp2,yk_rdp3}, so that both the density
of colored particles, and the interaction cross section, depend
nontrivially upon the Polyakov loop.

Consider first the density.  From Eq. (\ref{stat_dist_fncs}),
a background $Q$ field acts like an imaginary chemical potential
for color.  For ordinary statistical distribution functions, the only
way to obtain a small density of particles is if they are
heavy, and so Boltzmann suppressed.  A nonzero $Q$ field provides another.  
When $Q \neq 0$, the statistical distribution function 
for a given particle is complex
valued.  Thus while the distribution functions are
of order one in magnitude, because of cancelling phases, 
they can vanish after averaging over the $Q$'s. 
This happens in the confined phase at $\Nc = \infty$, where
the expectation value of $\tr \ploop^n$ vanishes for all $n$.
For a quark, which lies in the fundamental representation, the trace of the
statistical distribution function is given in Eq. (\ref{expansion}),
and vanishes term by term.  Just above $\Tc$, 
where $\ell$ is nonzero but small, the distribution function for such
a field vanishes as a single power of the loop,
\begin{equation}
\langle \sum_a n_{a} \rangle \sim \Nc\;  T^3 \; \ell \; ;
\label{fundamental_density}
\end{equation}
the bracket $\langle\cdots\rangle$ denotes the average over the 
$Q$ distribution, and an integral over the particles three momentum.

Gluons in the adjoint representation carry two $Q$ charges, 
so the Boltzmann expansion of the statistical distribution function
involves $\exp(i (Q^a - Q^b))$.  Summing over the color indices,
$\sum_{a} {\rm e}^{i Q^{a}}\sum_b {\rm e}^{- i Q^{b}} 
= \tr L \; \tr L^\dagger$,
so the density of gluons then vanishes as the square of the loop,
\begin{equation}
\langle \sum_{ab} n_{a b} \rangle \sim \Nc^{2}\; T^3 \; \ell^{2} \; ;
\label{gluon_density}
\end{equation}
we assume that the expectation of $\ell$ is real.
This suppression of the density has no analogy in the perturbative
regime, where the density of massless fields in necessarily a pure
number times $T^3$.

The average mean free path is the ratio of the density, divided by
a transport cross section.  For the scattering of two gluons,
\begin{equation}
\lambda^{gl}_\text{mfp}\sim
\frac{\langle \sum_{ab}n_{ab}\rangle }
{ \langle \sum_{ab,cd} \, n_{ab}\; n_{cd} \; \sigma_{ab;cd} \rangle } \; ;
\label{eq:averageMFP}
\end{equation}
$\sigma_{ab;cd}$ is the transport cross section for 
$2\rightarrow 2$ scattering between gluons with color indices $ab$ and $cd$.

For the sake of argument, assume first that there are no correlations
between different colors.  If true, then the cross section 
would factorize,
\begin{equation}
\langle \sum_{ab,cd}  n_{ab}\; n_{cd} \; \sigma_{ab;cd} \rangle =
\langle \sum_{ab} n_{ab}\rangle^2 \; \langle \sigma_{ab;cd} \rangle 
\sim \Nc^4 \; \sigma  \; \ell^4 \; .
\label{wrong_mean}
\end{equation}
Here $\sigma$ is a typical perturbative cross section,
$\sim g^4 \log(1/g)$.
The mean free path would then diverge at small $\ell$,
$\lambda^{gl}_\text{mfp}\sim 1/(\sigma \ell^2)$.
This factor of $1/\ell^2$ would cancel a similar factor in the density,
so that the shear viscosity would be as in ordinary perturbation theory,
$\eta \sim T/\sigma$, (approximately) independent of $\ell$
at small $\ell$.

As we shall see in the next section, however, there are strong correlations
between different colors, so that the 
cross section does not factorize as 
in Eq. (\ref{wrong_mean}).  Instead, it vanishes 
only quadratically with the loop,
\begin{equation}
\langle \sum_{ab,cd} n_{ab} \, n_{cd}\, \sigma_{ab;cd} \rangle \sim 
\Nc^{4}\; \sigma \; \ell^{2} \; ,
\label{density_semi}
\end{equation}
At first sight this dependence on $\ell$
seems counter intuitive: the scattering process involves
the scattering of two gluons, with color indices $a b$ and $c d$.  
Expanding the statistical distribution function 
for one gluon brings in powers of
$\exp(i (Q^a - Q^b))$, while expanding that for the other gluon brings in
powers of 
$\exp(i (Q^c - Q^d))$.  Summing over all color indices, the 
naive expectation is that the cross section
is proportional to the fourth power of the loop, one from each factor 
of $\exp(i Q)$.  
The scattering processes, however, include a planar diagram in which two
of the color indices are equal, say $b = c$.  For this
term, the factor of $\exp(- i Q^b)$ from one gluon
cancels, {\it identically},
against the factor of $\exp(+ i Q^c)$ from the other gluon.  
This explains why Eq. (\ref{density_semi}) is proportional to $\ell^2$,
and not to $\ell^4$, as in Eq. (\ref{wrong_mean}).

In the pure glue theory, then, from Eq. (\ref{eq:averageMFP}) the
mean free path is approximately independent of the expectation value of
the loop, $\lambda^{gl}_\text{mfp}\sim 1$.  
Even in the confined phase, the mass dimension of $Q$
is set by the temperature; thus the typical momentum remains
$\sim T$, and $\overline{p} \sim T$.
Consequently, from Eq. (\ref{eq:viscosityClassical})
the shear viscosity is proportional to the overall gluon density,
and so vanishes quadratically with the loop, $\eta \sim \ell^2$.

With dynamical quarks the counting differs, although
somewhat unexpectedly, the result is similar.  The
cross section for the scattering between a quark, with
color $a$, and an antiquark, with color $b$, is
\begin{equation}
\langle \sum_{a,b} n_{a} \, n_{b} \, \sigma_{a;b} \rangle  \; .
\label{density_semi_qk}
\end{equation}
If we could factorize the cross section, as in Eq. (\ref{wrong_mean}),
then this would be proportional to the density of quarks, squared, or
$\sim \ell^2$.  There is such a contribution when the colors are
unequal, $a \neq b$.  In addition, though, there is a planar diagram from
when the colors are equal, $a = b$.  
In this case the chemical potential for 
the quark, $\exp(i Q^a)$, cancels identically against that from
the antiquark, $\exp(- i Q^b)$, so that Eq. (\ref{density_semi_qk}) is
not $\sim \ell^2$, but constant.  
The mean free path for quarks,
\begin{equation}
\lambda^{qk}_\text{mfp}\sim
\frac{\langle \sum_{a}n_{a}\rangle }
{ \langle \sum_{a,b} \, n_{a}\; n_{b} \; \sigma_{a;b} \rangle } \; ;
\label{eq:averageMFPqk}
\end{equation}
is then proportional to the quark density, 
$\sim \ell$.  The average momentum remains $\sim T$, 
so that the shear viscosity is the product of the quark density, $\sim \ell$,
times the mean free path.  Thus like gluons, in all the quark
contribution to the shear viscosity is $\eta^{qk} \sim \ell^2$.   

Putting in the factors of the coupling constant, we find that either
in the pure glue theory, or in the theory with dynamical quarks, 
that the shear viscosity vanishes quadratically in the loop,
\begin{equation}
\eta \sim \frac{T^3}{g^4 \ln{1/g}} \; \ell^2 \; ,
\label{qualitiative_shear_viscosity}
\end{equation}
relative to the result in the perturbative QGP.  

We show later that the results for quarks is in fact numerically
close to that of the pure glue theory.  While we consider a large $\Nc$
limit in which $\Nf$ is also large, this similarity between the shear
viscosity in the pure glue theory, and with dynamical quarks, is rather
unexpected. 

\subsection{Derivation of the shear viscosity in the semi-QGP} 
\label{sec:derivation}
In this section, we use the Boltzmann equation to compute 
the shear viscosity in the semi-QGP
\cite{transport1,baym,transport2,amy,Asakawa,transport3,lebellac}.
Of course we assume that the coupling constant is small enough
so that a quasiparticle picture is applicable, and kinetic theory
is valid.  The computation is precisely like that in the ordinary 
QGP, except that we need to compute in the presence of a background
field for $A_0$.  In this section we consider a purely gluonic
theory, considering the case with dynamical quarks in the next
section.

If the statistical distribution function for a gluon with
color $a b$ is $f_{a b}$,
then in kinetic theory the energy-momentum tensor is
\begin{equation}
 T^{\mu\nu}
=2\int d\varGamma  \; P^{\mu}P^{\nu}f_{ab} \,.
\label{Tij}
\end{equation}
The factor of two in Eq.~(\ref{Tij}) comes from the following relativistic 
normalization of the phase space including color and spin:
\begin{equation}
\int d\varGamma=\sum_{ab,s}\int\frac{ d^3p}{(2\pi)^3 \; 2E} \, ,
\end{equation}
where $s$ denotes the spin of the gluon, and $E=|\bm{p}|$ is the energy of the gluon.
We are interested in a system 
slightly away from local thermal equilibrium, 
so we write the statistical distribution function as
\begin{equation}
f_{ab}=f_{ab}^{(0)}+\delta f_{ab} \; ,
\end{equation}
where $f_{a b}^{(0)}$ is the gluon distribution function in local thermal equilibrium, 
\begin{equation}
f^{(0)}_{ab}=\frac{1}{e^{\beta(u\cdot P^{ab})}-1} \,,
\label{eq:f0}
\end{equation}
with $P^{\mu;ab}=P^{\mu}-i\delta^{\mu 0}Q^{ab}$,
and $\delta f_{a b}$ measures the deviation from equilibrium.

The constant background field $\Acl$ induces an imaginary color charge.
We assume that after averaging over the $Q$'s, that
the net color charge vanishes.
A spatially dependent color charge arises in related
problems, such as for 't Hooft loops, when there is a background
color electric field~\cite{yk_rdp2,interface1,interface2,interface3}.

As discussed in Sec.~\ref{sec:effectiveTheory}, the background
field vanishes along the part of the contour in real time in 
Fig.~(\ref{timepath}).  Thus the time derivative in the Boltzmann
equation does not involve the background field,
\begin{equation}
2\, P^{\mu}\; \partial_{\mu}f_{a_1b_1}(\bx,\bp,t)=-C_{a_1b_1}[f]  \, .
\label{BoltzmannEq}
\end{equation}
We do not consider a force term, which may be dynamically generated
in an expanding plasma~\cite{Asakawa}.
In Eq.~(\ref{BoltzmannEq}),
$C[f]$ is the collision term for the scattering
of  $2\to2$ gluons,
\begin{equation}
\begin{split}
C_{a_1b_1}[f]=& \prod_{i=2}^{4}\int d\varGamma_{i} 
(2\pi)^4\delta^{4}(P_1+P_2-P_3-P_4)
|{\mathcal{M}}|^2\\
&\times \Bigl(f_{a_1b_1} f_{a_2b_2}(1+ f_{a_3b_3})(1+ f_{a_4b_4})
-(1+f_{a_1b_1})(1+ f_{a_2b_2}) f_{a_3b_3} f_{a_4b_4}\Bigl)\,,
\end{split}
\label{collision}
\end{equation}
where $\mathcal{M}$ is the scattering amplitude, which
depends upon the background field.
To leading order, it is necessary to include not only the scattering
of $2 \to 2$ particles, but also processes
involving splitting, of $1 \to 2$ particles,
and joining, for $2 \to 1$ particles \cite{baier,amy,AMY2}.
The processes of soft multiple splitting and joining are known as the 
Landau-Pomeranchuk-Migdal (LPM) effect~\cite{LPM}.
In this paper we compute to leading order in both the coupling, $g^2$,
and logarithms of $g$.  At leading logarithmic order, the result
is dominated by the scattering of $2 \to 2$ particles in the
$t$-channel, and the LPM effect can be neglected.
\begin{figure*}
\begin{center}
\includegraphics[width=160mm]{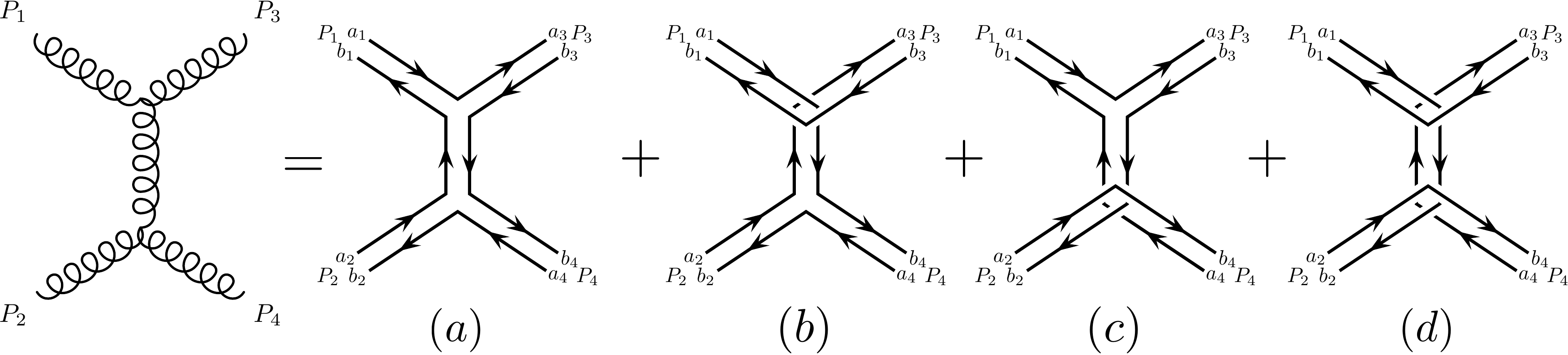}
\caption{Feynman diagram of the gluon scattering with the normal and double line notations in the $t$-channel.}
\label{fig:feynman_gluon}
\end{center}
\end{figure*}

In the pure glue theory, the corresponding diagram in the $t$-channel
is shown in Fig.~(\ref{fig:feynman_gluon}).  In the double line basis,
there are four diagrams.
The infrared singularity is cutoff by 
using a dressed propagator in the Hard Thermal Loop (HTL) approximation,
in the presence of the background field~\cite{htl,yk_rdp2},
\begin{equation}
D_{\mu\nu;ab,cd}(K) =  P^L_{\mu\nu} \; \frac{\bm{k}^2}{K^2} \;
D_{ab,cd}^L(K) + P^T_{\mu\nu}\; D_{ab,cd}^T(K)\,,
\label{propagator}
\end{equation}
$D^L$ and $D^T$ are the propagators 
for the longitudinal and transverse gluons in the HTL approximation,
respectively.  For the amplitude at tree level, gauge dependent
terms can be ignored.  Projection operators for the transverse
and longitudinal directions are
\begin{align}
P^T_{\mu\nu}=& 
g_{\mu i}\left(-g^{ij}-\frac{k^{i}k^{j}}{\bm{k}^{2}} \right)g_{i\nu} 
\,,\notag\\
P^L_{\mu\nu}=&- g_{\mu\nu}+\frac{K_\mu K_\nu}{K^2}- P^T_{\mu\nu} \,.
\end{align}
while the transverse and longitudinal propagators are
\begin{align}
D_{ab,cd}^T(K)=& \left[\frac{i}{K^2- m^2[x^2+\frac{x(1-x^2)}{2}
\ln\frac{x+1}{x-1}] } \right]_{ab,cd}\,,\notag\\
D_{ab,cd}^L(K)= &\left[\frac{i}{k^2+2m^2(1-\frac{x}{2}\ln\frac{x+1}{x-1}) } 
\right]_{ab,cd}\,,
\end{align}
where $x=k_0/|\bk|$. 
The thermal mass, $[m^2]_{ab,cd}$, 
depends upon the details of the color distribution~\cite{yk_rdp2}.
In general, it is not diagonal, and  
different components of $[m^2]_{aa, bb}$, with no summation over the
indices $a$ and $b$, mix with one another.
In the limit of large $\Nc$, 
$[m^2]_{ab,cd}$ becomes diagonal, 
and mixing is suppressed by $1/\Nc$. 
In the following, we work only to leading order of $\Nc$, in which
the propagators are diagonal:
\begin{align}
D_{ab,cd}^T(K)=&\;\delta_{ad}\; \delta_{bc}\; D_{ab}^T(K)\notag\\
=&\;\delta_{ad}\; \delta_{bc}\left[\frac{i}{K^2- [m^2]_{ab}
[x^2+\frac{x(1-x^2)}{2}\ln\frac{x+1}{x-1}] } \right],\notag\\
D_{ab,cd}^L(K)= &\; \delta_{ad}\; \delta_{bc}\; D_{ab}^L(K) \notag\\
=&\; \delta_{ad}\; \delta_{bc}
\left[\frac{i}{k^2+2[m^2]_{ab}(1-\frac{x}{2}\ln\frac{x+1}{x-1}) } \right]\; ,
\end{align}
where 
\begin{equation}
[m^{2}]_{ab}=\frac{g^{2}\Nc T^{2}}{6}\;
\frac{1}{2\Nc}\; 
\sum_{e=1}^{\Nc}(\mathcal{A}(Q^{ae})+\mathcal{A}(Q^{be})) \,,
\end{equation}
and 
\begin{equation}
\mathcal{A}(Q)= 1-6q(1-q) \,,
\end{equation}
where $q= Q/(2\pi T)$, and $q$ is defined modulo $1$~\cite{yk_rdp2}. 

We note that at large $\Nc$, the combination $g^{2}\Nc$, and so the Debye mass,
are of order one.  In absence of the background field, $Q=0$, 
$[m^2]_{ab}$ coincides with the ordinary
thermal mass, $[m^2]_{ab}=  g^2 \Nc T^2/6 $.

For the shear viscosity, we find that the result is proportional to
that in zero field, 
\begin{equation}
\eta =  \eta_\text{pert}\; \ratioR \;,
\label{eq_viscosity}
\end{equation}
where $\eta_\text{pert}$ is the 
usual result in perturbation theory \cite{amy},
\begin{equation}
\eta_\text{pert} = \frac{2 \Nc^{2} c_\eta \; T^3}{(g^2\Nc)^{2}\ln[1/(g^2\Nc)]} \, ;
\label{perturbative_shear_viscosity}
\end{equation}
$c_\eta$ is a constant of order one, which depends upon $\Nc$ and $\Nf$.

Note that at large $\Nc$, the free energy and so the entropy are large,
of order $\Nc^2$.  The shear viscosity is of the same order, so that
the ratio of the shear viscosity, to the entropy, is of order one.
Our task is now to derive $\ratioR$.

\subsubsection{Kinematics and Scattering amplitude}
\begin{figure}
\resizebox{0.45\textwidth}{!}{%
\includegraphics{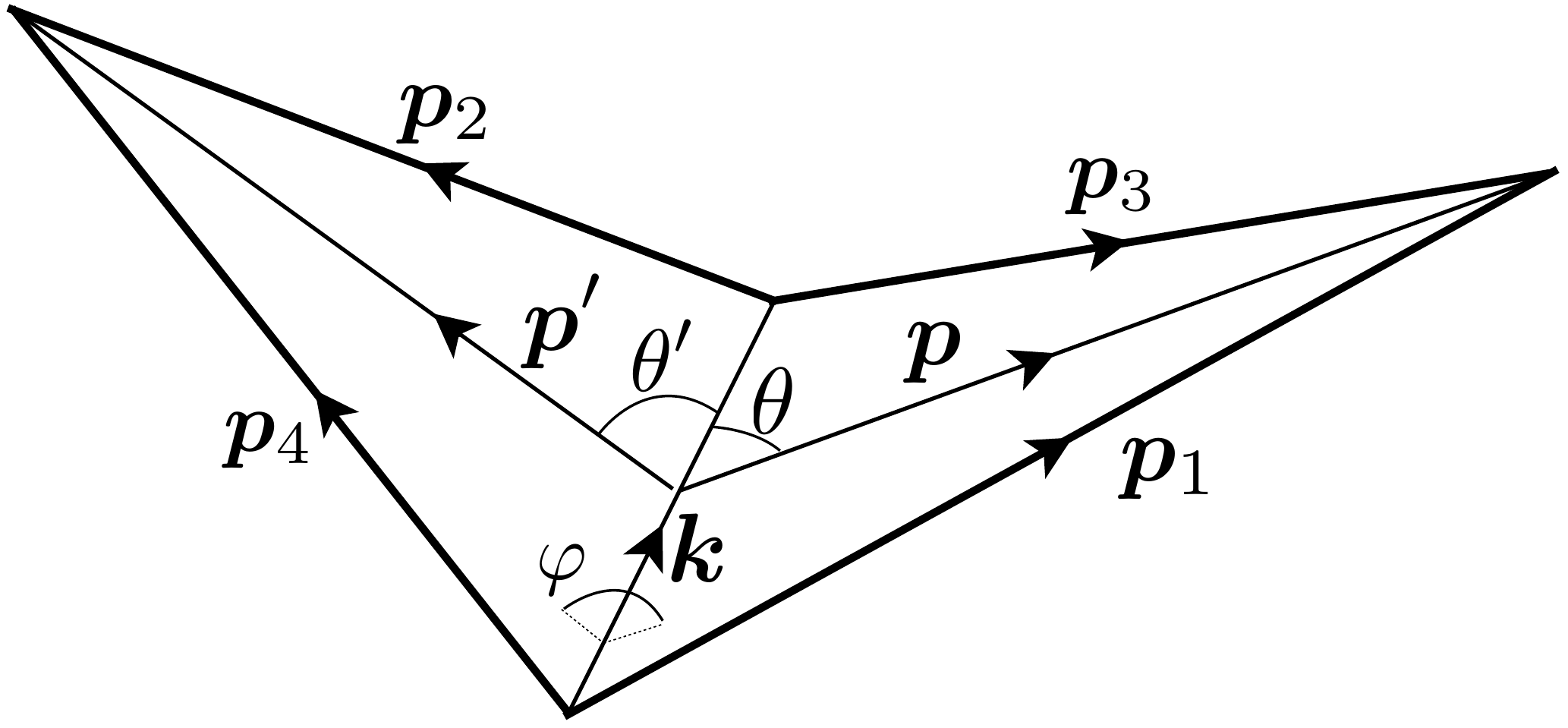}
}
\caption{Kinematics of the two body scattering.}
\label{fig:kinematics}
\end{figure}

Before solving the Boltzmann equation we review the
kinematics of scattering processes for $2 \to 2$ particles, involving
the exchange of a soft gluon ~\cite{baym}.
In terms of the Mandelstam variables,
\begin{equation}
s\equiv (P_1+P_2)^2\,,  \quad
t\equiv (P_1-P_3)^2 \,,  \quad
u\equiv (P_1-P_4)^2 \,.
\end{equation}
The dominant kinematic region in Fig.~(\ref{fig:feynman_gluon}) is 
$|t|\ll |s|\simeq |u|$, as then the amplitude, squared, is
enhanced by $\sim 1/t^2$.  We introduce the quantities:
\begin{equation}
P_1=P+\frac{K}{2} \,,\quad
P_2=P'-\frac{K}{2} \,,\quad
P_3=P-\frac{K}{2} \,, \quad
P_4=P'+\frac{K}{2} \,,
\end{equation}
in which the Mandelstam variables become
\begin{equation}
s= (P+P')^2 \,, \quad
t= K^2 \,, \quad
u= (P-P')^2  \,.
\end{equation}
There are three angles which enter, as shown in 
Fig.~(\ref{fig:kinematics}):
\begin{equation}
\cos\theta=\hat{\bk}\cdot \hat{\bp} \,, \quad
\cos\theta'=\hat{\bk}\cdot \hat{\bp'} \,,\quad
\cos\varphi =\frac{(\hat{\bp}\times\hat{\bk})
\cdot(\hat{\bp}'\times\hat{\bk})}
{|\hat{\bp}\times\hat{\bk}||\hat{\bp}'\times\hat{\bk}|} \,,
\end{equation}
where unit vector are written as $\hat{\bm{p}}=\bm{p}/|\bm{p}|$, {\it etc.}
For forward scattering, $\theta$ is close to $\theta'$, and
\begin{equation}
\cos\varphi
\simeq \frac{\hat{\bp}\cdot\hat{\bp}'-x^2}{1- x^2} \,,
\end{equation}
using $\cos\theta'\simeq x=k_0/|\bk|$. The Mandelstam variables become
\begin{align}
s\simeq&2pp'(1-x^2)(1-\cos\varphi) \,,\notag\\
u\simeq& -2pp'(1-x^2)(1-\cos\varphi)\,, \notag\\
t\simeq &-k^2(1-x^2)\,, 
\end{align}
with $p=|\bp|$, $p'=|\bp'|$, and $k=|\bk|$.
The amplitude in the $t$-channel is
\begin{equation}
i\mathcal{M}=J_{s_{1},s_{3}}^{\mu;ab}(P_1,P_3) 
\; D_{\mu\nu;ab,cd}(K) \; J_{s_{2},s_{4}}^{\nu;cd} (P_2,P_4) \,,
\end{equation}
where $J_{s,s'}^{\mu;ab}$ is the color current, 
and $D_{\mu\nu;ab,cd}$ is the gluon propagator in Eq.~(\ref{propagator}).
For forward scattering, $P_1\simeq P_3\simeq P$, 
and the form of the current is dictated by Lorentz symmetry,
\begin{equation} 
J_{s_{1},s_{3}}^{\mu;ab}(P_1,P_3)\simeq 2\, i\, g
\; P^\mu \; t^{ab}_r \; \delta_{s_{1},s_{3}} \,,
\end{equation}
where $t^{ab}_r$ is a generator of $\SU(N)$ in the representation $r$.
For the adjoint representation,
\begin{equation}
[t^{ab}_\text{adj}]^{cd,ef}= if^{(cd,ab,ef)} .
\end{equation}
The spin or helicity is denoted by $s_{1}$ and $s_{3}$.
The amplitude for gluon scattering becomes
\begin{equation}
\begin{split}
i\mathcal{M} =&-4\, g^2\,\delta_{s_1,s_3}\,\delta_{s_2,s_4}  
\sum_{ab,cd}\,t^{ab}_{13}\; t^{cd}_{24}\,  
P^{\mu}\,P^\nu D_{\mu\nu;ab,cd}(K) \\
=&-4\,  g^2\,
\delta_{s_1,s_3}\delta_{s_2,s_4}\,
\sum_{ab} \, t^{ab}_{13} \,t^{ba}_{24}\;  p\, p' 
\Bigl[D^{ab}_L(K)+(1-x^2)\cos\varphi \; D^{ab}_T(K)\Bigr] \,,
\end{split}
\end{equation}
using the shorthand notation
$t^{ab}_{ij}=[t_\text{adj}^{ab}]^{a_ib_i,a_jb_j}$.
The square of the amplitude is
\begin{equation}
\begin{split}
|{\mathcal{M}}|^2=&\sum_{ab,cd}16\,g^4\,p^2\,p'^2\, t_{13}^{ab}\;
t_{24}^{ba}\; t_{13}^{cd}\;t_{24}^{dc} \;\delta_{s_1,s_3}\; \delta_{s_2,s_4}\\
&\times\Bigl[D^{ab}_L(K) +(1-x^2)\cos\varphi \; D^{ab}_T(K)\Bigr] 
\Bigl[D^{cd}_L(K) +(1-x^2)\cos\varphi \; D^{cd}_T(K)\Bigr]^* \,.
\end{split}
\label{eq:squaredAmplitude}
\end{equation}

In bare perturbation theory, the thermal mass vanishes, and
this becomes
\begin{equation}
|{\mathcal{M}}|^2
=4\, g^4 \; \sum_{ab,cd} t_{13}^{ab}\;t_{24}^{ba} \;t_{13}^{cd}\;
t_{24}^{dc} \;\delta_{s_1,s_3} \;\delta_{s_2,s_4}  \frac{s^2}{t^2}  \, .
\end{equation}

It is rather useful to 
write phase space in terms of the variables
$k$, $p$, $p'$, $x$ and $\varphi$, instead of $\bm{p}_{1}$ to $\bm{p}_{4}$.
By energy-momentum conservation,
\begin{equation}
\begin{split}
\delta(E_1+E_2-E_3-E_4)\simeq& \delta\bigl(k(\cos\theta-\cos\theta')\bigr) \\
=&\frac{1}{k}\delta(\cos\theta-\cos\theta') \,,
\end{split}
\end{equation}
the integral over phase space becomes
\begin{equation}
\begin{split}
&\prod_{i=2}^{4}\int  d\varGamma_{i}(2\pi)^4
\delta^{(4)}(P_1+P_2-P_3-P_4) \\
&\qquad\simeq  \frac{1}{2^2(2\pi)^6} 
\prod_{i=2}^{4} \sum_{a_{i}b_{i},s_{i}}
\int d k \, k \; d p\, p\; d p' \, d x \;d \varphi \;
(2\pi)^3\delta^{(3)}\bigl(\bp_1\!-\!\bp\!-\!\frac{\bk}{2}\bigr) \,.
\end{split}
\end{equation}

\subsubsection{Chapman-Enskog method}
Because the Boltzmann equation is nonlinear, it is not easy to
solve exactly.  Instead, we use the 
Chapman-Enskog method to estimate the transport coefficients~\cite{chapman}.
Start with a solution of the Boltzmann equation 
local thermal equilibrium $f_{ab}^{(0)}$, Eq.~(\ref{eq:f0}).
By the conservation of energy-momentum and charge, the collision term
vanishes, and gives the following relation:
\begin{equation}
\frac{f^{(0)}_{a_1b_1}}{1+f^{(0)}_{a_1b_2}}
\frac{f^{(0)}_{a_2b_2}}{1+f^{(0)}_{a_2b_2}}=
\frac{f^{(0)}_{a_3b_3}}{1+f^{(0)}_{a_3b_3}}
\frac{f^{(0)}_{a_4b_4}}{1+f^{(0)}_{a_4b_4}} \,.
\label{EMC}
\end{equation}
We are interested in a nonequilibrium system which is close
to local thermal equilibrium, and so expand
the distribution function around $f_{ab}^{(0)}$,
\begin{equation}
f_{ab}=f_{ab}^{(0)}+\epsilon \, f_{ab}^{(1)}+\epsilon^{2}
\, f_{ab}^{(2)}+\cdots \, .
\end{equation}
This corresponds to expanding in spatial derivatives, or equivalently,
the Knudsen number.  The parameter $\epsilon$
keeps track of the order of derivatives, and is taken to one after
computation.
$f^{(n)}$ gives the transport coefficients of $n$-th order, 
{\it e.g.}, $f^{(1)}$ gives the shear and bulk viscosities.
The collision term is expanded in powers of $\epsilon$ as
\begin{equation}
C[f] = \sum_{n=0}^{\infty} \epsilon^{n} C^{(n)} \, ,
\end{equation}
where
\begin{equation}
C^{(n)} =\frac{1}{n!}\left.\frac{d^{n}}{d\varepsilon^{n}} 
\; C[f]\right|_{\epsilon=0} \,.
\end{equation}
The first term, $C^{(0)}$, 
obviously vanishes, because $f^{(0)}_{ab}$ is the solution in 
thermal equilibrium.
In the left hand side of the Boltzmann equation, 
the spatial derivative, $\partial_{i}$, is taken to be of order $\epsilon$. 
In the Chapman-Enskog method, 
the time derivative is expanded in powers of $\epsilon$,
\begin{equation}
\partial_{t}=\sum_{n=1}^{\infty}\epsilon^{n}\partial_{t}^{(n)} \,.
\end{equation}
We show in the following that
$\partial_{t}^{(n)}$ is determined by energy-momentum conservation 
and local thermal equilibrium,  
Eqs.~(\ref{eq:localthermalequilibrium}) and (\ref{eq:conservationlow}).
$f^{(n)}_{ab}$ is parametrized as
\begin{equation}
f^{(n)}_{ab}=f^{(0)}_{ab}(1+f^{(0)}_{ab})\; \varPhi^{(n)} 
\end{equation}
for useful.
To $\epsilon^n$, the Boltzmann equation is written as
\begin{equation}
2\, P^{\mu}\; \partial_{\mu}^{(n)}\, f_{a_1b_1} =
-\collisionOperator \; \varPhi^{(n)}_{a_1b_1} +K^{(n)}_{a_1b_1} \,,
\label{eq:linearBolzmannEq}
\end{equation}
where the partial derivative, to order $\epsilon^n$, is
\begin{equation}
P^{\mu}\; \partial_{\mu}^{(n)}f_{ab} =
p^{0}\sum_{k=1}^{n}\partial_{t}^{(k)}f^{(n-k)}_{ab}+
p^{i}\, \partial_{i}\, f^{(n-1)}_{ab} \,,
\end{equation}
and
\begin{equation}
K^{(n)}_{a_1b_1}= -C^{(n)}_{a_1b_1} + \collisionOperator\; \varPhi^{(n)}_{a_1b_1} \,.
\end{equation}
We introduce the linearized operator $\collisionOperator$,
\begin{equation}
\begin{split}
\collisionOperator \varPhi^{(n)}_{a_1b_1}  =&
 \prod_{i=2}^4\int d\varGamma_{i} \;
(2\pi)^4\delta^{(4)}(P_1+P_2-P_3-P_4) \; |{\mathcal{M}}|^2 \\
&\quad\times
  f^{(0)}_{a_1b_1}\; f^{(0)}_{a_2b_2}\; \left(1+f^{(0)}_{a_3b_3}\right)
\left(1+f^{(0)}_{a_4b_4}\right) 
\left(\varPhi^{(n)}_1+\varPhi^{(n)}_2
-\varPhi^{(n)}_3-\varPhi^{(n)}_4 \right) \,.
\end{split}
\end{equation}
In vanishing background field, 
the operator $\collisionOperator$ is self-adjoint and positive semidefinite:
\begin{equation}
\int d\varGamma \; \varPhi^\dagger \left(\collisionOperator \; \varPhi \right) 
= \int d\varGamma ( \collisionOperator\; \varPhi)^\dagger \varPhi \; \geq 0 
\qquad\text{for}\quad \varPhi\neq0\,.
 \label{eq:linearOperator}
\end{equation}
While positivity is not automatic in a nonzero background field, we assume
it is, after averaging over color.  Our analysis shows no signs of 
such unphysical behavior.

Equation (\ref{eq:linearBolzmannEq}) is a 
linear equation 
for $\varPhi^{(n)}$.  Since $K_n$ and $\partial_\mu^{(n)}f_{ab}$ are 
as functionals of $f_{ab}^{(k)}$ 
for $0\leq k \leq n-1$, and does not include $f_{ab}^{(n)}$, 
$\varPhi^{(n)}$ can be obtained order by order. 

There is an ambiguity, however.  
$\collisionOperator$ contains zero eigenvalues, 
which correspond to conserved currents, $\varPhi=P^\mu$ and $1$.
To remove this ambiguity,
we require that at each order in $\epsilon$,
the deviation from thermal equilibrium is orthogonal 
to the zero modes,
\begin{equation}
\int d \varGamma \; P^{0}\; P^{\mu} \; f^{(n)}_{ab} = 0 
\end{equation}
for $n\geq 1$.  Then the conserved charge density only depends upon
$f^{(0)}_{ab}$,
\begin{equation}
T^{0\mu}= 2\int d\varGamma \; P^{0}\; P^{\mu} \, f^{(0)}_{ab} \,.
\label{eq:localthermalequilibrium}
\end{equation}
We also require that energy and momentum 
are conserved order by order in $\epsilon$,
\begin{equation}
2\, \partial_{t}^{(n)} \int d\varGamma  P^{0}P^{\nu}f^{(0)}_{ab} +
2\, \partial_{i} \int d\varGamma  P^{i}P^{\nu}f^{(n-1)}_{ab} =0 \,,
\label{eq:conservationlow}
\end{equation}
which defines the time derivative to $\epsilon^n$.

The viscosities are determined at leading order,
\begin{equation}
2\, P^{\mu}\; \partial_{\mu}^{(1)}f_{a_1b_1} = -\collisionOperator\; \varPhi^{(1)}_{a_1b_1} \,,
\label{eq:leadingOrder}
\end{equation}
after using $K_{1}=0$.  The left hand side is
\begin{equation}
\begin{split}
2\, P^{\mu}\; \partial_{\mu}^{(1)}\; f_{ab} \!= \!&
-\!2\; P^{ab;\nu} \; P^{\mu} \;
f^{(0)}_{ab}
\left(1+f^{(0)}_{ab}\right) \left( u_{\nu}\;
\partial_{\mu}^{(1)} \beta  +\beta \; \partial_{\mu}^{(1)} u_{\nu}\right) \,.
\end{split}
\end{equation}
Working in the local rest frame, where
$u_{0}=1$, $u_{i}=0$, $\partial^{(1)}_{\mu}u_{0} = 0$ and $\partial^{(1)}_{\mu}u_{i}\neq 0$, 
\begin{equation}
\begin{split}
2\, P^{\mu}\, \partial_{\mu}^{(1)}f_{ab} 
=-f^{(0)}_{ab}(1+f^{(0)}_{ab})\Big[
2E^{2}&\Big(\partial^{(1)}_{t}\beta +\frac{\beta}{3}\partial_{i}u_{i}\Big) 
+ 2Ep^{i}(\partial_{i}\beta +\beta \partial_{t}^{(1)}u_{i})\\
& +\beta p^{i}p^{j}\sigma_{ij} -iQ^{ab}  (E\partial^{(1)}_{t}\beta+p^{i}\partial_{i}\beta)
  \Big] \,.
  \end{split}
\label{eq:BoltzmannEquationII}
\end{equation}
By assumption, the last term, proportional to $Q^{a b}$,
vanishes after averaging over color, and so we drop it in
computing the shear viscosity.
This equation contains $\partial_{t}^{(1)}$, which is yet
to be determined. 
From 
Eq.~(\ref{eq:conservationlow}), to leading order
\begin{equation}
\int d\varGamma \; P^{\nu} P^{\mu}\; \partial_{\mu}^{(1)}f^{(0)}_{ab} =0 \,,
\end{equation}
which reads
\begin{align}
\partial_{t}^{(1)}\beta+\frac{\beta}{3}\; \partial_{i}u_{i}&=0  \,, \notag\\
\beta\; \partial_{t}^{(1)}u_{i}+\partial_{i}\beta&=0 \,.
\label{eq:timeDerivative}
\end{align}
Using Eq.~(\ref{eq:timeDerivative}) in Eq.~(\ref{eq:BoltzmannEquationII}),
only last term on the left hand side survives,
\begin{equation}
2\; P^{\mu}\, \partial_{\mu}^{(1)}f 
=-\beta \, f^{(0)}_{ab}\left(1+f^{(0)}_{ab}\right)\, p^{i} p^{j}\, \sigma_{ij} \,.
\end{equation}
The trace of this equation vanishes, so that the bulk viscosity vanishes
in this calculation.  At linear order, the Boltzmann equation becomes
\begin{equation}
\beta f^{(0)}_{a_1b_1}\left(1+f^{(0)}_{a_1b_1}\right)\, p^{i}p^{j}\,\sigma_{ij}
=\collisionOperator\;\varPhi^{(1)}_{a_1b_1} \,.
\label{eq:firstBoltzmann}
\end{equation}
Acting with the 
inverse of the collision operator, $\collisionOperator^{-1}$,
on both sides of Eq.~(\ref{eq:firstBoltzmann}),
we obtain the formal solution,
\begin{equation}
\varPhi^{(1)}=  \collisionOperator^{-1}\,\beta\, f^{(0)}_{a_1b_1}\left(1+f^{(0)}_{a_1b_1}\right) 
\, p^{i}p^{j} \,\sigma_{ij} \,.
\end{equation}
Note that the condition of orthogonality, 
Eq.~(\ref{eq:localthermalequilibrium}),
ensures that zero modes are projected out, so that
the inverse operator $\collisionOperator^{-1}$ is well-defined.
We parametrize $\varPhi^{(1)}$ as
\begin{equation}
\varPhi^{(1)}=\left(p^{i}p^{j}-\frac{p^{2}}{3}\delta^{ij}\right)
\sigma_{ij} \; \chi(p)=\sigma_{ij}\; \chi^{ij}(p) \, .
\label{eq:solution1}
\end{equation}
Remember that $\sigma_{i j}$ is traceless, so that here we can add
$p^{2}\delta^{ij}\sigma_{ij}/3$ on the left hand side,
and take $\chi^{i j}$ to be traceless.
The transport coefficients are independent of one another, so we write
\begin{equation}
\beta f_{a_1b_1}^{(0)}\left(1+f_{a_1b_1}^{(0)}\right)
\left(p^{i}p^{j}-\frac{1}{3}\delta^{ij}p^{2}\right)=
\collisionOperator\chi^{ij}_{a_1b_1} \,.
\label{eq:BoltzmanEquationB}
\end{equation}
Using the solution of the Boltzmann equation, we
rewrite $\varPi^{ij}$ in Eq.~(\ref{eq:Tmunu})
to obtain
\begin{equation}
\varPi^{(1)ij}=2\int d\varGamma \; p^{i}p^{j}\; f^{(1)}_{ab}
= 2\int d\varGamma \; p^{i}p^{j}\; f_{ab}^{(0)}
\left(1+f_{ab}^{(0)}\right)\varPhi^{(1)}
\,,
\end{equation}
where $\varPi^{(1)ij}$ denotes $\epsilon^1$.
Using Eq.~(\ref{eq:solution1}), and performing the integral over angles
gives
\begin{equation}
\int \frac{d\Omega}{4\pi} \; p^{i}p^{j}p^{k}p^{l}
=\frac{1}{15}\; p^{4}
\left(\delta^{ij}\delta^{kl}
+\delta^{ik}\delta^{jl}+\delta^{il}\delta^{jk}\right) \,,
\label{eq:angularIntegral}
\end{equation}
we find
\begin{equation}
\varPi^{(1)ij}
= \frac{2}{5}\;
\sigma^{ij}\int d\varGamma \left(p^{k}p^{l}-\frac{1}{3}\delta^{kl}p^{2}\right)
f^{(0)}\left(1+f^{(0)}\right)\chi^{kl}  \,.
\end{equation}
The rank tensor two in the integrand is nothing but 
$\collisionOperator\chi^{kl}$ in Eq.~(\ref{eq:BoltzmanEquationB}),
so that
\begin{equation}
\varPi^{(1)ij}= 
\sigma^{ij}\; \frac{2T}{5}\int d\varGamma \;
\chi^{kl} \; \collisionOperator\; \chi^{kl}\, .
\end{equation}
The shear viscosity is then
\begin{equation}
\eta = \frac{2T}{5}\int d\varGamma \; 
\left( \chi^{ij}\, \collisionOperator\, \chi^{ij} \right) \,.
\label{eq:viscosity1}
\end{equation}
In order to determine $\chi_{ij}$, 
we need to know the inverse of a linear operator,
with an infinite number of matrix elements. 

We first find a formal solution, expanding $\chi$ in orthogonal polynomials,
$\chi=\sum c_{n} \chi_{n}(p)$, and 
\begin{equation}
\beta^{6}\int d\varGamma \;
f_{ab} \left(1+f_{ab}\right)
\; p^{4} \; \chi_{n}(p)\ \chi_{m}(p) = d_{n}\; \delta_{nm} \,,
\label{eq:orthogonalCondition}
\end{equation}
where $d_{n}$ is a normalization constant. 
We choose $\chi_{n}=\sum_{m=0}^{n}b_{nm}(p/T)^{m}$ to be a monic polynomial, 
{\it i.e.}, the leading coefficient is unity, $b_{nn}=1$.  This polynomial 
can be systematically obtained by Gram-Schmidt orthogonalization.
The first three terms are
\begin{align}
\chi_{0}=&1 \,, \\
\chi_{1}=&\left(\frac{p}{T}\right)+b_{10} \,,\\
\chi_{2}=&\left(\frac{p}{T}\right)^{2}+b_{21}\left(\frac{p}{T}\right)+b_{20} \,,
\label{eq:chi2}
\end{align}
where the $b_{nm}$ are dimensionless constants:
\begin{align}
b_{10}&=  -\; \frac{h_{1}}{h_{0}}\,,\\
b_{20}&=\frac{h_{1}h_{3}-h_{2}^{2}}{h_{0}h_{2}-h_{1}^{2}}\,,\\
b_{21}&=\frac{h_{1}h_{2}-h_{0}h_{3}}{h_{0}h_{2}-h_{1}^{2}}\,,
\end{align}
where
\begin{equation}
\begin{split}
h_{n}=&
\;\beta^{6-n}\int d\varGamma \; f_{ab}\left(1+f_{ab}\right)
\; p^{4+n} \\
          =&\;\frac{\varGamma(6+n)}{2\pi^{2}}
\; \sum_{k=1}^{\infty}\;\frac{|\tr \, \ploop^{k}|^{2}}{k^{5+n}} \,.\\
\end{split}
\end{equation}
The normalization constants are 
obtained from Eq.~(\ref{eq:orthogonalCondition})--(\ref{eq:chi2}) as
\begin{align}
d_{0}=& \; h_{0} \,,\\
d_{1}=& \; h_{2}+h_{1}b_{10}  \,,\\
d_{2}=& \; h_{4}+h_{3}b_{21} + h_{2}b_{20}   \,.
\end{align}
Taking $\int d\varGamma \, \chi^{ij}_{n}$ to Eq.~(\ref{eq:BoltzmanEquationB}), 
the Boltzmann equation reduces to a matrix form,
\begin{equation}
S_{n} = \sum_{m}\;\collisionOperator_{nm}\;c_{m} \,,
\end{equation}
where
\begin{equation}
S_{n}=\beta\int d\varGamma \;
\chi_{n}^{ij} \; f^{(0)}_{ab}\left(1+f^{(0)}_{ab}\right)
\left(p^{i}p^{j}-\frac{p^{2}}{3}\delta^{ij}\right)  \,,
\end{equation}
with
\begin{equation}
\collisionOperator_{nm}=\int d\varGamma \;
\left( \chi_{n}^{ij}\,\collisionOperator\,\chi_{m}^{ij} \right) \, .
\end{equation}
Formally, the solution can be written as
\begin{equation}
\chi^{ij}=\sum_{n,m}\chi_{n}^{ij}\; [\collisionOperator^{-1}]_{nm}\; S_{m} \,.
\label{eq:solution2}
\end{equation}
Inserting this into Eq.~(\ref{eq:viscosity1}), we obtain 
\begin{equation}
\eta = \frac{2\, T}{5}\; \sum_{n,m}S_{n}
\left[\collisionOperator^{-1}\right]_{nm}
S_{m} \,.
\label{eq:Viscosity2}
\end{equation}
From Eqs.~(\ref{eq:angularIntegral}) and (\ref{eq:orthogonalCondition}),
\begin{equation}
S_{n} = \frac{2\, T^{5}}{3}\; d_{0}\; \delta_{n,0} \,,
\end{equation}
from which the shear viscosity is
\begin{equation}
\eta = \frac{8\, T^{11}}{45}\; d_{0}^{2}\,
\left[\collisionOperator^{-1}\right]_{00} \,.
\end{equation}
This involves the zero-zero component of the inverse of $\collisionOperator$.
In general, finding the inverse of $\collisionOperator$ is not easy,
as it involves an infinite number of matrix elements.
It is thus necessary to truncate the number of matrix elements.
This gives a lower bound on the solution,
because Eq.~(\ref{eq:Viscosity2}) has a quadratic form, 
and by Eq.~(\ref{eq:linearOperator}), $\collisionOperator$ is positive semidefinite.
We will see the solution converges quickly
as the size of the matrix for $\collisionOperator$ increases, with the
solution for one matrix element within
$0.6\%$ of the exact solution 
(see Fig.~(\ref{fig:ndependence}) and related
discussion following Eq.~(\ref{eq_pert_shear_viscosity})).
Using just one matrix element, we obtain
an approximate form for the shear viscosity,
\begin{equation}
\eta \approx \frac{8\, T^{11}}{45}\; \frac{d_{0}^{2}}{\collisionOperator_{00}} \,.
\label{eq:Viscosity3}
\end{equation}

We now turn to estimating the matrix element,
\begin{equation}
\begin{split}
\collisionOperator_{nm} =&
\frac{1}{4} \prod_{i=1}^4\int d\varGamma_{i} \;
(2\pi)^4\delta^{(4)}\left(P_1+P_2-P_3-P_4\right) \; |{\mathcal{M}}|^2 \;
  f^{(0)}_{a_1b_1}\; f^{(0)}_{a_2b_2}
\left(1+f^{(0)}_{a_3b_3}\right)\left(1+f^{(0)}_{a_4b_4}\right) \\
&\qquad\times  \left(\chi_{n}^{ij}(p_{1})+\chi_{n}^{ij}(p_{2})
-\chi_{n}^{ij}(p_{3})-\chi_{n}^{ij}(p_{4})\right) \,
\left(\chi_{m}^{ij}(p_{1})+\chi_{m}^{ij}(p_{2})
-\chi_{m}^{ij}(p_{3})-\chi_{m}^{ij}(p_{4})\right) \,.
\end{split}
\label{eq:matrixelement}
\end{equation}
The coefficient $1/4$ is caused by symmetrizing the collision term.
We are interested only in the forward scattering amplitude,
where the exchanged momentum, $k$, is small.
In the integrand of Eq.~(\ref{eq:matrixelement}),
to leading order in $k$,
\begin{equation}
\begin{split}
\chi_{n}^{ij}(p_{1})-\chi_{n}^{ij}(p_{3})
&=\chi_{n}^{ij}\left(p+\frac{k}{2}\right)
-\chi_{n}^{ij}\left(p-\frac{k}{2}\right)\\
&\quad=\left(p^{i}k^{j}+k^{i}p^{j}
-2(\bp\cdot \bk)\frac{p^{i}p^{j}}{p^{2}}\right)\chi_{n}(p)\\
&\qquad+\left(p^{i}p^{j}-\frac{p^{2}\delta^{ij}}{3}\right)
(\bp\cdot \bk)\, \frac{1}{p^{3}}\, \left(p^{2}\chi_{n}(p)\right)' 
+\mathcal{O}(k^{2}) \,.
\end{split}
\end{equation}
Further,
\begin{equation}
\begin{split}
& \left(\chi_{n}^{ij}(p_{1})+\chi_{n}^{ij}(p_{2})
-\chi_{n}^{ij}(p_{3})-\chi_{n}^{ij}(p_{4})\right) 
\left(\chi_{m}^{ij}(p_{1})+\chi_{m}^{ij}(p_{2})
-\chi_{m}^{ij}(p_{3})-\chi_{m}^{ij}(p_{4})\right) \\
&\simeq2\left(p^{2}k^{2}-(\bm{p}\cdot\bm{k})^{2}\right)
\chi_{n}(p)\chi_{m}(p) 
+\frac{2}{3}\left(\bm{p}\cdot\bm{k}\right)^{2}
\; \frac{1}{p^{2}}\; \left(p^{2}\chi_{n}(p)\right)' (p^{2}\chi_{m}(p))'
+(p\leftrightarrow p') \\
&=k^{2}\Bigl(2(1-x^{2})\; p^{2}\; \chi_{n}(p)\chi_{m}(p)
+\frac{2x^{2}}{3} \; \left(p^{2}\chi_{n}(p)\right)' 
\left(p^{2}\chi_{m}(p)\right)'\Bigr) +(p\leftrightarrow p')\\
&\equiv k^{2}F_{nm}(p,p',x) \, .
\end{split}
\end{equation}
In the second line, we dropped the cross term $(\chi_{n}^{ij}(p_{1})-\chi_{n}^{ij}(p_{3}))(\chi_{m}^{ij}(p_{2})-\chi_{m}^{ij}(p_{4}))$
that vanishes in the integral of $\theta$ and $\phi$.
The function $F_{nm}(p,p',x)$ is independent of $k$;
the matrix element becomes
\begin{equation}
\begin{split}
\collisionOperator_{nm} &\simeq
\frac{1}{4} \; \prod_{i=1}^4\int d\varGamma_{i} \;
(2\pi)^4\delta^{(4)}(P_1+P_2-P_3-P_4) 
\; |{\mathcal{M}}|^2 \\
&\qquad\times  f^{(0)}_{a_1b_1}\; f^{(0)}_{a_2b_2}\;
\left(1+f^{(0)}_{a_3b_3}\right)\left(1+f^{(0)}_{a_4b_4}\right)  \;
  k^{2}\; F_{nm}(p,p',x)\,.
\end{split}
\label{eq:ff(1+f)(1+f)}
\end{equation}
Expanding the statistical distribution functions at small $k_0$,
\begin{equation}
\begin{split}
f^{(0)}_{a_1b_1} \; f^{(0)}_{a_2b_2} \;
\left(1+f^{(0)}_{a_3b_3}\right)\left(1+ f^{(0)}_{a_4b_4}\right)
=&\sum_{n_1=1}^\infty\sum_{n_2=1}^\infty\sum_{n_3=0}^\infty
\sum_{n_4=0}^\infty \exp{\Big[-\beta\sum_{j=1}^4n_j(E_j-iQ^{a_jb_j})\Big]}\\
\simeq&\sum_{n_1=1}^\infty\sum_{n_2=1}^\infty
\sum_{n_3=0}^\infty\sum_{n_4=0}^\infty\mathcal{G}(Q,n)\\
&\quad \times {\rm e}^{-\beta(n_1(p+\frac{k_0}{2})+
n_2(p-\frac{k_0}{2})+n_3(p'+\frac{k_0}{2})+n_4(p'-\frac{k_0}{2}))} \,,
\end{split}
\end{equation}
where
\begin{equation}
\mathcal{G}(Q,n)=\exp \Big[i\beta\sum_{j=1}^4 n_jQ^{a_jb_j} \Big] \, .
\end{equation}
Note $\mathcal{G}(0,n) = 1$.
Equation~(\ref{eq:ff(1+f)(1+f)}) has no singularity at $k_{0}=0$, so
taking $k_0 \to 0$,
\begin{equation}
\begin{split}
&f^{(0)}_{a_1b_1} \; f^{(0)}_{a_2b_2} 
\left(1+f^{(0)}_{a_3b_3}\right)\left(1+ f^{(0)}_{a_4b_4}\right)
\simeq\sum_{n_1=1}^\infty\sum_{n_2=1}^\infty\sum_{n_3=0}^\infty\sum_{n_4=0}^\infty \mathcal{G}(Q,n) 
e^{-\beta((n_1+n_3)p+(n_2+n_4)p')} \,.
\label{eq:distributionfunction}
\end{split}
\end{equation}
Writing the collision term using this, and using the
explicit form of the squared amplitude in Eq.~(\ref{eq:squaredAmplitude}) as
\begin{equation}
\begin{split}
\collisionOperator_{nm}=&\; 2\; \frac{g^4}{(2\pi)^5}
\prod_{i=1}^4\sum_{a_ib_i}\sum_{n_1=1}^\infty
\sum_{n_2=1}^\infty\sum_{n_3=0}^\infty\sum_{n_4=0}^\infty\mathcal{G}(Q,n) 
\; t_{13}^{ab}\, t_{24}^{ba}\, t_{13}^{cd}\, t_{24}^{dc}
\int_0^\infty d p\; p^2 \; e^{-\beta p(n_1+n_3)} \\
&\quad\times\int_0^\infty d p'p'^2 e^{-\beta p'(n_2+n_4)}
\int_{-1}^1 d x\int_{-\pi}^{\pi}\frac{d \varphi}{2\pi} 
\int_0^\infty d k \; k^3\; F_{nm}(p,p';x) \\
&\qquad\quad \times\Bigl[D^{ab}_L(K) +(1-x^2)\cos\varphi \; D^{ab}_T(K)\Bigr] 
\Bigl[D^{cd}_L(K) +(1-x^2)\cos\varphi \; D^{cd}_T(K)\Bigr]^* \,.
\end{split}
\label{eq:matrixElement2}
\end{equation}
The infrared singularity for the longitudinal component is cut off
by the Debye mass, but that for the transverse component is regularized
by the dynamical screening mass, $\sim x m^2$.  The is infrared safe
for $x \neq 0$, and is finite after integration over $x$.
Consider the following integral:
\begin{equation}
\begin{split}
\int_0^T d k \; \frac{k^3}{(k^2+m_1^2)(k^2+m_2^2)}
&\approx\frac{1}{4}\ln\left(\frac{T^4}{m_1^2\, m_2^2}\right) \\
&\approx \frac{1}{2}\ln\frac{1}{g^2\Nc} \,,  
\end{split}
\end{equation}
where we use $m_1^2$ and $m_2^2 \sim g^2\Nc T^2$, and we keep only
the logarithmic term.  This expression is valid as long
as $\tr\, \ploop \neq 0$.  Under these approximations, the integral
over $k$ in Eq.~(\ref{eq:matrixElement2}) can be done, giving
\begin{equation}
\begin{split}
\int_0^\infty d k \; k^3 \; |D_L+(1-x^2)\cos\varphi \; D_T|^2
\approx  \frac{1}{2}(1-\cos\varphi)^2\; \ln\frac{1}{g^2\Nc}   \,.
\end{split}
\end{equation}
We can then perform the angular integral,
\begin{equation}
\begin{split}
&\int_{-1}^1 d x \int_{-\pi}^{\pi}\frac{d \varphi}{2\pi} \;
F_{nm}(p,p',x) \left(1-\cos\varphi\right)^2 \\
&=\;4\, p^{2}\; \chi_{n}(p)\chi_{m}(p)
+\frac{2}{3} \; \left(p^{2}\chi_{n}(p)\right)'\left(p^{2}\chi_{m}(p)\right)'
+\left(p\!\leftrightarrow \!p'\right)\\
&=\frac{20}{3}\; p^{2} \; \sum_{l=0}^{n+m} 
C_{nm,l} \left[\left(\frac{p}{T}\right)^{l}
+\left(\frac{p'}{T}\right)^{l}\right]
\,,
\label{eq:integF}
\end{split}
\end{equation}
where 
\begin{equation}
C_{nm,l}=\sum_{s=0}^{n}\sum_{t=0}^{m}\; b_{n,s}b_{m,t} \;
\left(1+\frac{1}{10}\; (st+2l)\right)\delta_{s+t,l} \,.
\end{equation}
This integral is symmetric under exchange of $p$ and $p'$, 
so we can replace $p^{2}+p'^{2}$ by $2p'^{2}$.  Using
\begin{equation}
\int_0^\infty d p \; p^2 \; e^{-\beta p(n_1+n_3)}
\int_0^\infty d p'\; p'^{4+l}\;\beta^{l}e^{-\beta p'(n_2+n_4)} 
=\frac{48\, T^8\; \varGamma(5+l)}{(n_1+n_3)^3(n_2+n_4)^{5+l}\varGamma(5)} \,,
\end{equation}
the matrix element $\collisionOperator_{nm}$ becomes
\begin{equation}
\begin{split}
\collisionOperator_{nm}=&
\frac{20\,T^8 \,g^4}{\pi^5}
\ln\left(\frac{1}{g^2\Nc}\right)\prod_{i=1}^4 \; \sum_{a_i,b_i} \;
\sum_{n_1=1}^\infty\; \sum_{n_2=1}^\infty\;
\sum_{n_3=0}^\infty\; \sum_{n_4=0}^\infty \\
&\times\mathcal{G}(Q,n) \; 
t_{13}^{ab}\; t_{24}^{ba} \; t_{13}^{cd} \; t_{24}^{dc} \;
\frac{1}{(n_1+n_3)^3}
\sum_{l=0}^{n+m}\frac{C_{nm,l}}{(n_2+n_4)^{5+l}}\;
\frac{\varGamma(5+l)}{\varGamma(5)} \,.
\end{split}
\end{equation}
This matrix element is $\sim T^{8}g^{4}\ln(1/g^2\Nc)$, 
so that $\eta\sim T^{3}/(g^{4}\ln1/(g^2\Nc))$, as expected.
Thus in the semi-QGP, the factors of temperature and the coupling
constant are the same as in the perturbative QGP; the 
dependence upon the background field appears only through
$\mathcal{G}(Q,n)$.

The next step is to sum over color. 
There are sixteen diagrams contributing to the viscosity.
Interference terms are not planar diagrams, and so can be dropped at
large $\Nc$.  In Fig~({\ref{fig:feynman_gluon}),
the contribution of diagrams $(a)$ and $(b)$ coincides with 
that of $(d)$ and $(c)$. The color structure of diagrams $(a)$ and $(b)$ are 
\begin{align}
\mathcal{G}_{(a)}(Q,\!n)& =
\exp\bigl[i\beta\bigl(Q^{a}(n_1+n_3)+Q^{b}(n_2-n_1)
+Q^{c}(n_4-n_3)-Q^{d}(n_2+n_4)\bigr) \bigr] \,,\notag\\
\mathcal{G}_{(b)}(Q,n)& = 
\exp\bigl[i\beta\bigl(Q^{a}(n_1+n_4)+Q^{b}(n_2+n_3)
-Q^{c}(n_1+n_3)-Q^{d}(n_2+n_4)\bigl)\bigl] \,,
\end{align}
respectively. Introducing
\begin{equation}
m_1=n_1+n_3\,, \quad m_2 =n_2+n_4\,,\quad
m_3=n_3\,, \quad m_4 =n_4 \,,
\end{equation}
the matrix element becomes
\begin{equation}
\begin{split}
\collisionOperator_{nm}=&\frac{10\, T^8\, (g^2\Nc)^2 \Nc^2}{\pi^5}
\ln \left(\frac{1}{g^2\Nc}\right)
\sum_{m_1=1}^\infty\sum_{m_2=1}^\infty
\sum_{m_3=1}^{m_1}\sum_{m_4=1}^{m_2} \sum_{l=0}^{n+m}
\frac{\ell_{m_1}}{m_1^3}\;
\frac{\ell_{m_2}}{m_2^{5+l}}\; \frac{\varGamma(5+l)}{\varGamma(5)}
\; C_{nm,l}
\\
&\quad\times (
\ell_{m_4-m_3}\ell_{m_2+m_3-m_1-m_4} 
+\ell_{m_1-m_3+m_4}\ell_{m_2-m_4+m_3}) \,.
\end{split}
\end{equation}
This is the final form for $\collisionOperator_{nm}$.
If the background vanishes, 
\begin{equation}
\begin{split}
\collisionOperator_{nm}=&\frac{20\, T^8\, (g^2\Nc)^2 \Nc^2}{\pi^5}
\; \zeta(2)\; \zeta(4)\; 
\ln \left(\frac{1}{g^2\Nc}\right)
\sum_{l=0}^{n+m}\;
\frac{\zeta(4+l)}{\zeta(4)}\; \frac{\varGamma(5+l)}{\varGamma(5)}
\; C_{nm,l} \, ,
\end{split}
\end{equation}
where $\zeta(z)$ is the Riemann zeta-function.
In particular, for the inverse of $\collisionOperator$,
the zero-zero component dominates,
\begin{equation}
\collisionOperator_{00}=\frac{20\, T^8\, (g^2\Nc)^2 \Nc^2}{\pi^5}
\; \zeta(2)\; \zeta(4)\; \ln \left(\frac{1}{g^2\Nc}\right) \, ,
\end{equation}
where $\zeta(2)=\pi^2/6$ and $\zeta(4)=\pi^4/90$.
At finite $\Nc$, $\Nc^{4}$ is replaced by $\Nc^{2}(\Nc^{2}-1)$.
In zero background field,
\begin{equation}
d_{0}(\ell=1)=\frac{60 \, \Nc^{2}}{\pi^{2}}\; \zeta(5) \, ,
\end{equation}
where $\Nc^{2}$ is replaced by $(\Nc^{2}-1)$ at finite $\Nc$.

In zero background field, at large $\Nc$ the shear viscosity is
\begin{equation}
\begin{split}
\eta_{\text{pert}}&= \frac{8\, T^{11}}{45}\;
\frac{d_{0}^{2}}{\collisionOperator_{00}(\ell=1)} \\
&=540\; \zeta^{2}(5)
\; \Bigl(\frac{2}{\pi}\Bigr)^{5}\frac{T^{3}\Nc^2}{(g^2 \Nc)^2 
\ln1/(g^2\Nc)} \,.
\end{split}
\label{eq_pert_shear_viscosity}
\end{equation}
This result was first 
obtained by Baym {\it et al.} \cite{baym}.
Our result is valid at large $\Nc$, with the correct result at
finite $\Nc$ is obtained by replacing the overall factor of $\Nc^2$ 
in the numerator by $\Nc^2 -1 $.

Comparing with Eq.~(\ref{perturbative_shear_viscosity}),
the coefficient $c_\eta=26.98$. This result is very close to the exact 
result, $c_{\eta}=27.13$, which was obtained numerically 
by Arnold, Moore, and Yaffe \cite{amy}.  We confirmed their
results with several large matrices.  
In Fig~(\ref{fig:ndependence}) we show the dependence of
$[\collisionOperator^{-1}]_{00}(\ell=0)$ upon the size of the matrix.
The solution converges quickly as the size of the matrix increases,
with the solution for one matrix element within $\sim 0.6\%$ of the
exact result.  Therefore, in the following we limit ourselves to
the approximation of one matrix element.
\begin{figure}
\resizebox{0.45\textwidth}{!}{\includegraphics{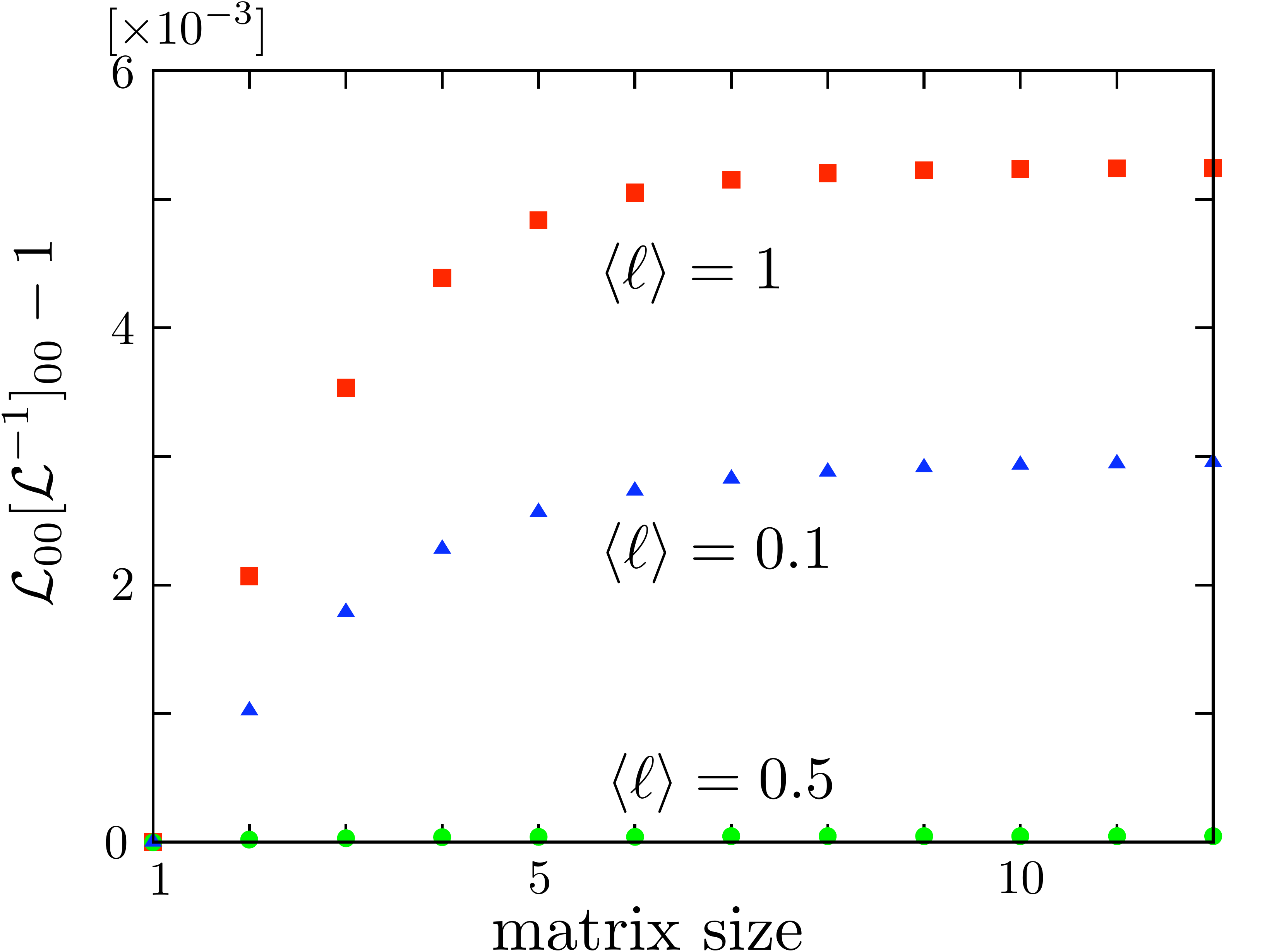}}
\caption{Matrix size dependence of $\collisionOperator_{00}[\collisionOperator^{-1}]_{00}-1$.}
\label{fig:ndependence}
\end{figure}

In order to compare the viscosity in the semi-QGP to $\eta_\text{pert}$, we calculate the ratio $\ratioR$ in Eq.~(\ref{eq_viscosity}),
\begin{equation}
\ratioR=\frac{\eta}{\eta_\text{pert}} 
=\frac{d_{0}^{2}(\ell=1)[\collisionOperator^{-1}(\ell=1)]_{00}}{d_{0}^{2}(\ell)[\collisionOperator^{-1}(\ell)]_{00}}
\,.
\end{equation}
Inserting the explicit form, we obtain 
\begin{equation}
\begin{split}
\ratioR =&\left[\frac{1}{\zeta(5)}\sum_{n=1}^\infty \frac{|\ell_n|^2}{n^5}\right]^2 
\Biggl[\frac{1}{2\zeta(4)\zeta(2)}\sum_{m_1=1}^\infty\sum_{m_2=1}^\infty
\sum_{m_3=1}^{m_1}\sum_{m_4=1}^{m_2}\frac{\ell_{m_1}}{m_1^3}\frac{\ell_{m_2}}{m_2^5}\\
&\quad \times (
\ell_{m_4-m_3}\ell_{m_2+m_3-m_1-m_4}
+\ell_{m_1-m_3+m_4}\ell_{m_2-m_4+m_3})  \Biggr]^{-1}\,.
\end{split}
\label{eq:ratio}
\end{equation}
This is our main result.  If the background field vanishes, 
{\it i.e.}, all $\ell_n=1$,  this ratio is unity, $\ratioR=1$.
Conversely, if all $\ell_n=0$, this ratio vanishes, $\ratioR=0$.
Numerically we will see how 
$\ratioR$ changes as a function of $\ell$
in the next section.
Before going to the numerics, 
we estimate $\collisionOperator$ near $\Tc$, where
the Polyakov loop is small. Neglecting higher powers of $\ell$, 
and higher moments, $\ell_{n}$ for $n\geq 2$, 
\begin{equation}
\ratioR
=
\frac{\pi^{6}}{270\,\zeta^{2}(5)}
\; \frac{\ell^{2} }{1+\ell^{2}} \simeq 3.31 \frac{\ell^{2} }{1+\ell^{2}}\,.
\label{eq:viscosityAtSmallL}
\end{equation}
Thus the shear viscosity is $\sim \ell^2$ at small $\ell$.

Let us see why the shear viscosity becomes small.
Comparing Eqs.~(\ref{eq:viscosityClassical}) and 
(\ref{eq:averageMFP}) to Eq.~(\ref{eq:Viscosity3}),
one identifies
\begin{equation}
\sum_{ab } n_{ab }  \sim T^{3}d_{0}(\ell) \;\; ; \;\;
\sum_{{ab,cd}} n_{ab} \, n_{cd}\, \sigma_{ab;cd} \sim 
\frac{\mathcal{L}_{00}(\ell)}{T^{4}}
\; .
\end{equation}
As mentioned in Sec.~\ref{sec:shearViscosityKineticTheory},
both the gluon number $d_{0}\sim \Nc^{2}\ell^{2}$, and the 
transport cross section multiplied by density squared.
$\sum_{{ab,cd}} n_{ab} \, n_{cd}\, \sigma_{ab;cd}\sim \Nc^{4}\ell^{2}$,
are small when $\ell$ is.

This suggests two important points.
First, gluons (and other colored fields)
are suppressed as $T \rightarrow T_c^+$.
This ``bleaching'' of colored fields is most natural
in the confined phase; it is universal, and depends only upon
the color representation, independent of the fields spin, flavor, or mass.
Secondly, even if the coupling
constant is moderate, the correlation between gluons in the 
semi-QGP effectively becomes large; this is why the viscosity
is small when $\ell$ is.
Conversely, the interaction between small color representations
becomes large: 
{\it e.g.}, quark and antiquark scattering
includes two channels, an adjoint an a singlet.
For small $\ell$, the singlet channel dominates that in the adjoint.
This reflects that as one approaches the confined phase, 
interactions are dominated by the formation of color singlet states,
such as glueballs and mesons.
\begin{figure}
\resizebox{0.65\textwidth}{!}{%
\includegraphics{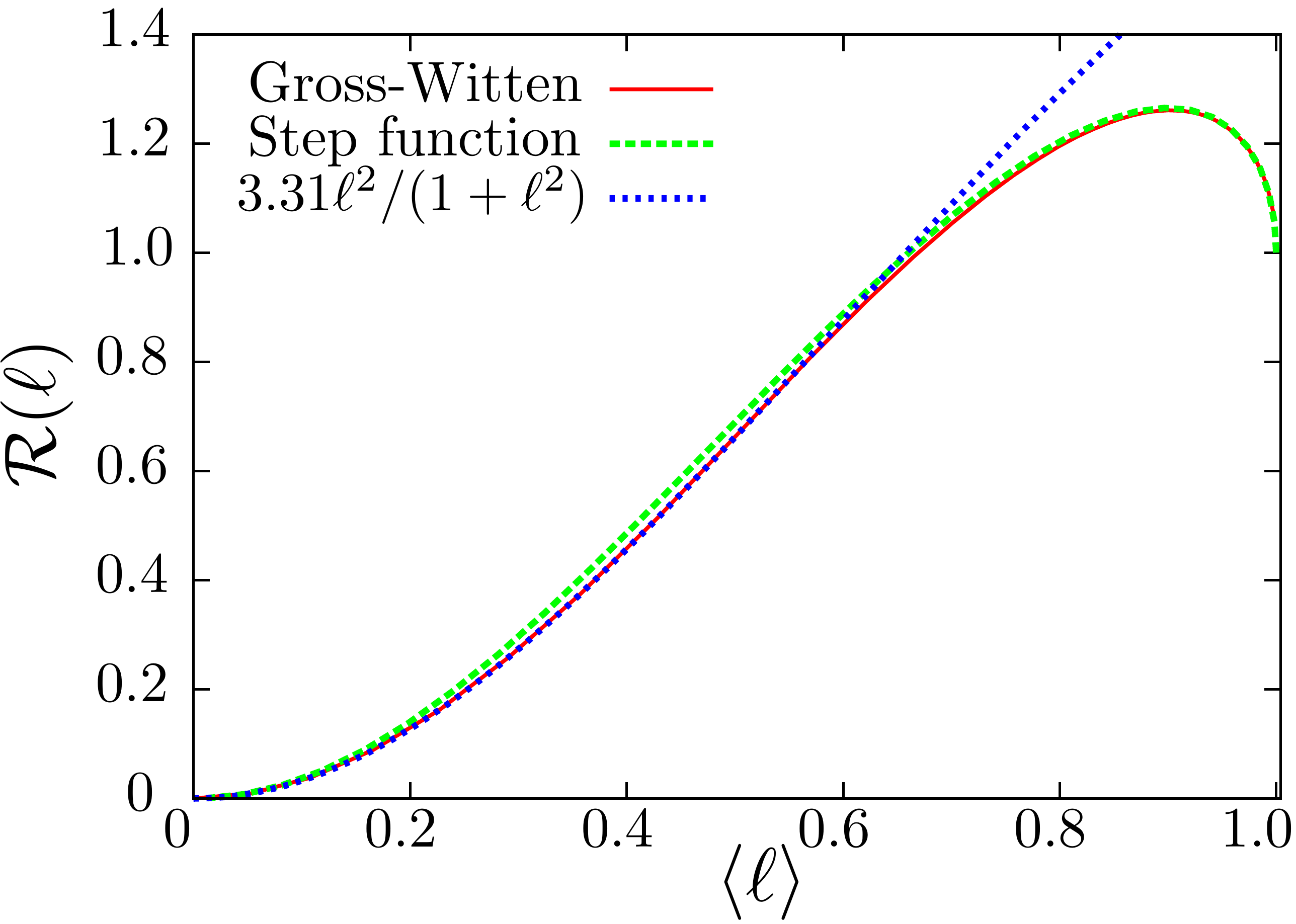}
}
\caption{Numerical results for $\ratioR$, Eq.~(\ref{eq:ratio}).
The two ansatzes for the eigenvalues are a step function (dashed line)
and Gross-Witten (solid line).  The dotted line denotes a simple
analytic approximation at small $\ell$, Eq.~(\ref{eq:viscosityAtSmallL}). }
\label{fig:gluon}
\end{figure}

\section{Numerical Results}\label{sec:NumericalResults}
\subsection{The shear viscosity in the semi-QGP}
In this section we compute numerically the ratio 
$\ratioR$ in Eq.~(\ref{eq:ratio}).  This is the
ratio of the shear viscosity in the semi-QGP, when $\langle\ell_{n}\rangle \neq 1$,
to that in the perturbative QGP, when all $\langle \ell_{n}\rangle = 1$.
The results are plotted in Fig.~(\ref{fig:gluon}) as a
function of $\langle \ell\rangle$.  Of course
the ratio $\ratioR$ is a function not only of the 
first moment of the eigenvalue distribution,
$\langle\ell\rangle$, but of higher moments,
$\langle \ell_{n}\rangle$. 
The problem is that while $\langle \ell \rangle$ is easily extracted
from numerical simulations on the lattice, the full eigenvalue distribution
is not.  

To illustrate the range of possible results, as 
mentioned in Sec.~\ref{sec:effectiveTheory} we consider
two different distributions, which we hope are representative.
The first is a step function, Eq.~(\ref{eq:distributionStep});
the second is that of a Gross-Witten model, Eq.~(\ref{eq:distributionGW}).
The different results for $\ratioR$ is illustrated in
Fig.~(\ref{fig:gluon}): there the result for
a step function is shown by a dashed green line, 
and that for a Gross-Witten model, a solid red line.
The difference between the two distributions are at most
a few percent, over the entire range of $\ell$.  In fact, up to $\ell=0.6$, 
an approximate solution without higher moments 
in Eq.~(\ref{eq:viscosityAtSmallL}), denoted by
a blue dotted line shown in Fig.~(\ref{fig:gluon}), 
is close to the other two distributions.
This suggests that at least for the shear viscosity, the results
are not very sensitive to higher moments of the eigenvalue distribution.
We do not know if this is generic, or special to the shear viscosity,
computed at leading logarithmic order

We find suppression of the shear viscosity at small $\ell$,
$\ratioR\sim \ell^2$ as $\ell \rightarrow 0$.  
Conversely, we find that
$\ratioR$ is enhanced for $\ell \sim 1.0$.
For $\ell\simeq0.9$; $\ratioR\sim 1.25$; near 
$\ell = 1$, an approximate
form is $\sim 1+1.47\sqrt{1-\ell}$. 
We do not have a simple explanation for this enhancement; if it
is a peculiarity of working at leading logarithmic order, {\it etc.}

\begin{figure}
\resizebox{.80\textwidth}{!}{%
\includegraphics{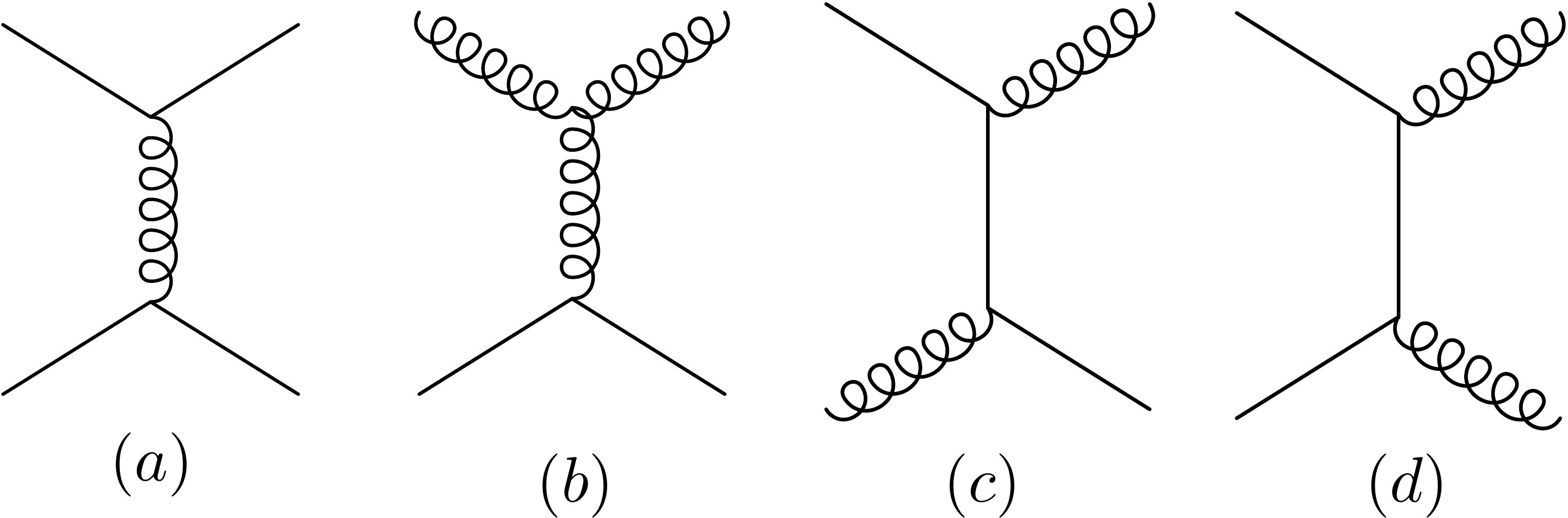}}
\caption{Feynman diagram of quark contributions to the shear viscosity in the leading order. The straight line denotes quarks or antiquarks, and the curly line denotes gluons.}
\label{fig:quarkScatteringT-channe}
\end{figure}

We also computed the shear viscosity with dynamical quarks;
details will be presented elsewhere \cite{hidaka}.
As in the pure glue theory, the dominant processes to leading
logarithmic order are those involving the exchange of a soft particle
in the $t$-channel.  The diagrams which contribute are illustrated in
Fig.~(\ref{fig:quarkScatteringT-channe}).  There are contributions from the
scattering between a quark and an antiquark, or scattering between
two quarks, Fig.~(\ref{fig:quarkScatteringT-channe}.a);
scattering between a gluon and a quark, 
Fig.~(\ref{fig:quarkScatteringT-channe}.b);
and Compton scattering between quarks and gluons,
Fig.~(\ref{fig:quarkScatteringT-channe}.c) and 
Fig.~(\ref{fig:quarkScatteringT-channe}.d).

However, the physical processes which dominate 
are very different from the pure glue theory.  At small $\ell$,
the dominant scattering processes are not between two gluons,
but between a quark and an antiquark, as shown in
Fig.~(\ref{fig:quarkScatteringT-channe}.a).
Contributions to the shear viscosity arise from two parts,
$d_{0}$ and $\collisionOperator_{00}$, shown in Eq.~(\ref{eq:Viscosity3}).
$d_{0}$ is proportional to the number density; for gluons,
this is of order $\ell^{2}$, while with quarks, it is
$\sim \ell$.  In $\collisionOperator_{00}$, gluon-gluon scattering
gives $\collisionOperator_{00}\sim \ell^2$; the scattering of a quark
and an antiquark includes a channel,
Fig.~(\ref{fig:quarkScatteringT-channe}.a), which is color singlet, 
so that $\collisionOperator_{00} \sim \ell^0$.  
Therefore, for a purely gluonic theory, the shear viscosity is
$\eta \sim d_0^2/\collisionOperator_{00}\sim (\ell^2)^2/\ell^2 \sim \ell^2$;
with dynamical quarks, it is also
$\eta \sim d_0^2/\collisionOperator_{00}\sim (\ell)^2/\ell^0 \sim \ell^2$.
The similar behavior of $\eta$ at small $\ell$ must presumably reflect
some more fundamental physics, although we do not know yet what it is.

In the numerical results, 
we took the ratio of the number of flavors to color to be fixed
at several values, equal to
$\Nf/\Nc=0$ (the pure glue theory), $1/3$, $2/3$ and $1$.
The results are illustrated in Fig.~(\ref{fig:quarks_Viscos}) \cite{yk_rdp1}. 
There is a weak, but non-negligible, dependence upon $\Nf/\Nc$.

\begin{figure}
\resizebox{0.65\textwidth}{!}{%
\includegraphics{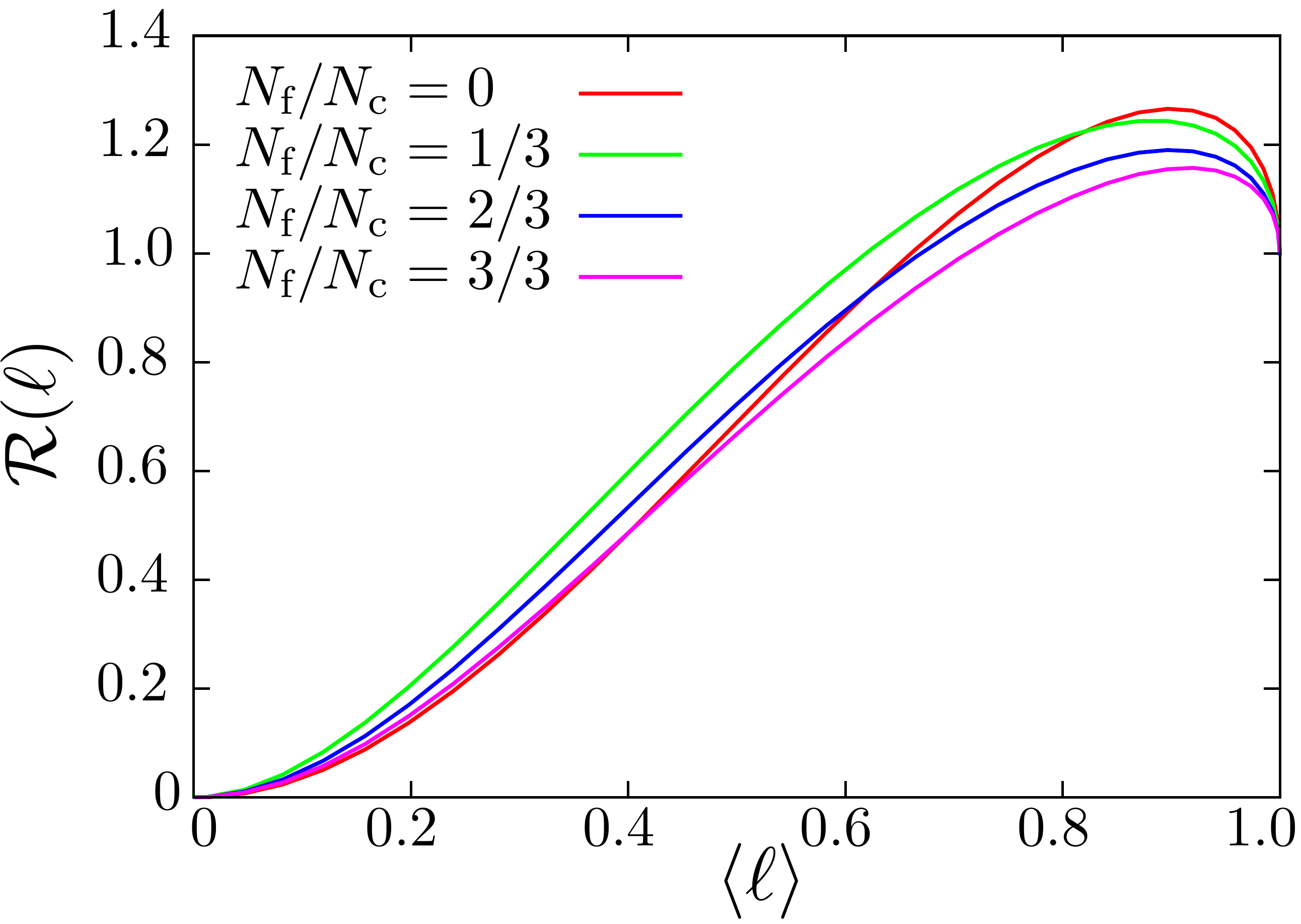}}
\caption{Flavor dependence of the viscosity.}
\label{fig:quarks_Viscos}
\end{figure}

\subsection{The ratio of the shear viscosity to the entropy} 
\label{sec:eta/s}

In applications to hydrodynamics, what matters is not the shear
viscosity by itself, but the ratio of the shear viscosity to the entropy
density, $\eta/s$.  For ${\cal N}=4$ supersymmetric gauge theories,
the gauge/gravity duality gives
$\eta/s = 1/4\pi$ \cite{susy1,susy2}, 
which is conjectured to be a universal lower bound.

In this subsection we compute the behavior of $\eta/s$ in the 
semi-QGP.  In principle, with a complete theory of the semi-QGP 
the entropy density would follow from the eigenvalue density.  
Since such an eigenvalue density is not presently known,
we use results directly from the lattice, $\slat$~\cite{cheng}.
This is not a significant limitation, since any complete theory
of the semi-QGP would necessarily give an entropy density in agreement
with $\slat$.

We also need to know the temperature dependence of the coupling
constant in QCD.  We use the formula for the running of the coupling
to two loop order,
\begin{equation}
\Nc \, \alpha_s = \frac{\Nc \, g^{2}(T)}{4 \pi} \approx
\frac{4\pi}{\beta_{0}(x)\ln\frac{T^{2}}{\varLambda^{2}}
+\frac{\beta_{1}(x)}{\beta_{0}(x)}\ln\ln\frac{T^{2}}{\varLambda^{2}}},
\end{equation}
where the coefficients of the beta function at large $\Nc$ are 
\begin{align}
\beta_{0}=&\frac{1}{3}(11-2x) \, ,\\
\beta_{1}=&\frac{1}{3}(34-13x)\, , 
\end{align}
with $x=\Nf/\Nc$.
We determine the renormalization mass scale, $\varLambda$,
such that $\alpha_s(T_c) \approx 0.3$ \cite{coupling}.
This value of the coupling is moderate in strength, so that our
computation, to leading logarithmic order, may not be reliable.
To parametrize this uncertainty, we introduce
a phenomenological parameter $\kappa$ to represent effects at higher
order in the shear viscosity,
\begin{equation}
\eta = 
\frac{2 \Nc^{2}\, c_\eta\;\ratioR \; T^3}
{(\Nc g^2(T))^{2}\, \ln(\kappa/(g^2(T)\Nc))} 
\,.
\end{equation}

\begin{figure}
\resizebox{0.65\textwidth}{!}{%
  \includegraphics{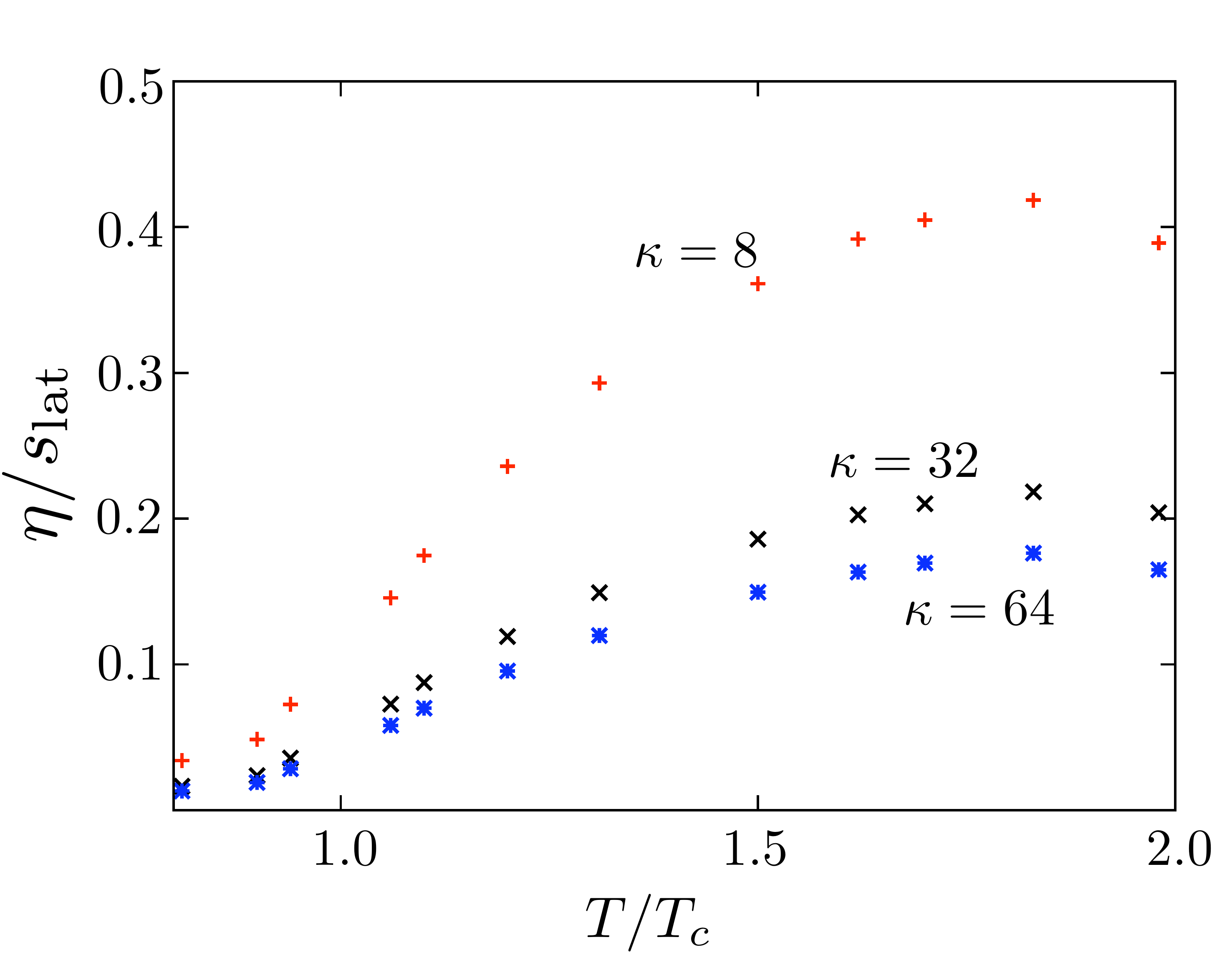} }
\caption{Ratio of the shear visosity to the lattice entropy.}
\label{fig:etaoversLat}
\end{figure}

In Fig.~(\ref{fig:etaoversLat}) we plot $\eta/\slat$,
when $\Nc=\Nf$, for three values of $\kappa$, $\kappa=8$, $32$, and $64$.
As the temperature decreases, the coupling increases, and the
shear viscosity, $\eta \sim 1/g^4(T)$, decreases.  For larger values
of $\kappa$, the dependence on the temperature is weaker.  

The shear viscosity 
$\eta  \sim \Nc^2/(g^2 \Nc)^2$; with $g^2 \Nc$ fixed at large
$\Nc$, $\eta \sim \Nc^2$.
Our computation is valid to leading
logarithmic order, and includes the effects of quarks and gluons.  
Hadrons, such as mesons and glueballs, are subleading, and contribute
$\sim 1$ to $\eta$.
Thus when at small $\ell \sim 1/\Nc$, our approximations break down.  
We expect that this is the reason why our results for
$\eta/\slat$ violates the conjectured lower bound, near and below $\Tc$.

\section{``Bleaching'' color in the semi quark gluon plasma}
\label{bleaching}

We conclude with some general comments about our results, and some
speculative remarks.

The simplest way of viewing deconfinement in a non-Abelian gauge theory
is in analogy to ionization in an Abelian plasma
\cite{yk_rdp1,ichimaru}.   
Define the lower limit of the semi-QGP as $\mTsemi$,
and the upper limit, as $\pTsemi$.  
Recall that in QCD, 
lattice simulations indicate that the 
semi-QGP exists between $\mTsemi\approx 0.8 \, \Tc$ and 
$\pTsemi\approx 3-4\, \Tc$,
as discussed in the Introduction and Refs.
\cite{yk_rdp1,yk_rdp2,yk_rdp3}.

There is no ionization of color
below $\mTsemi$, so the only states are colorless hadrons.
The ionization of color is partial in the semi-QGP, between
$\mTsemi$ and $\pTsemi$, and total above
$\pTsemi$, in the perturbative QGP.  A simple corollary of this picture is
that scattering process involving colored states disappears
as $T \rightarrow \mTsemi$ from above.  One may say that the scattering
of colored particles is ``bleached'' by the vanishing of the Polyakov loop.
In this paper we computed how the shear viscosity is bleached,
an example especially relevant for experiment \cite{hydro}.

One result which followed from our analysis, but which we did not
stress, is the following.  The Polyakov loop is
related to the propagator of an infinitely heavy, test quark.  Yet we
find that the bleaching of dynamical, light colored fields, such as gluons
and massless quarks, is identical to that of heavy quarks.  That is,
at least within the approximations in which we work,
the bleaching of color is universal, {\it independent} of the mass of
the dynamical fields.  

Thus it is necessary to discuss what assumptions are implicit in our
analysis; these are equivalent to the statement that
the semi-QGP exists only in a relatively narrow window
about $\Tc$.  

That $\mTsemi$ is not too far below $\Tc$ implies that
it is reasonable to speak of confinement in a theory
with dynamical quarks.  It is not difficult to think of gauge theories for
which this is not true.  Consider a gauge theory
with three colors and eight massless flavors, where lattice simulations
appear to indicate that chiral symmetry is spontaneously broken in the
vacuum \cite{lattice_largeNf}.  If so, then there is a chiral phase
transition at which this symmetry is restored at a finite, nonzero
temperature.  Nevertheless, in this theory the chiral phase transition
bears little relation to the deconfining phase transition
of the pure gauge theory.  This follows simply by counting the number
of degrees of freedom: the hadronic pressure, from an ideal gas of 
$63$ Goldstone bosons, is large, $\approx 2/3$ that of the ideal QGP.
Concomitantly, presumably the expectation value of the 
(renormalized, fundamental representation) Polyakov
loop is also
significant at temperatures well below that for the restoration of
chiral symmetry.  In this case, one might say that the semi-QGP begins
at a temperature far below $\Tc$, but the physics is really just
eight flavors of quarks throughly wash out 
the $Z(3)$ global symmetry of the pure glue $\SU(3)$ theory.  
The physics is dominated by chiral symmetry, and its restoration,
with deconfinement a minor perturbation on that.

The other assumption is that $\pTsemi$ is not much greater
than $\Tc$.  Assume that the contrary were true, $\pTsemi \gg \Tc$:
then we expect that
both the ratio of the pressure, to the ideal gas pressure,
and also the (renormalized, fundamental representation) 
Polyakov loop would deviate from unity, from
$\pTsemi$ all of the way down to $\Tc$.
While straight Polyakov loops presumably dominate near $\pTsemi$,
it is unreasonable to expect that they dominate all of the way down
to $\Tc$, if $\pTsemi \gg \Tc$.  Instead,
we suggest that in this case it is necessary to include
thermal Wilson lines, and the corresponding Polyakov loops, which
oscillate an even number of times in $\tau : 0 \rightarrow 1/T$.
(The number of oscillations must be even
because a Wilson loop, as a quantity formed from bosonic fields, must
be periodic in $\tau$.)  As one goes to temperatures much lower
than $\pTsemi$, the effects of oscillatory Polyakov loops become
more significant.  An effective theory of Polyakov loops can still
be constructed; it is just that a new type of Polyakov loops 
have to be folded in.

If effects from oscillatory Polyakov loops are important, then it is
clear that
the propagation of light fields {\it would} differ from that of heavy
fields.  A heavy field propagates in a straight line in imaginary time,
but a light field performs a random walk; 
the lighter the field, the more the dominant
paths in the path integral include those which fluctuate from a straight path.
Thus if $\pTsemi$ were much larger than $\Tc$,
the light fields would feel the effects of oscillatory Wilson
lines, but the heavier fields would not.  

In fact, even in a pure gauge theory, where the
semi-QGP exists only in a narrow region in temperature,
there is no reason why oscillatory Polyakov loops could not be
constructed and measured.  On the lattice, it will be awkward to
discretize them, but they could be measured, on a sufficiently fine lattice.
Our principal 
assumption is then that they are not needed when the semi-QGP exists
in a narrow region in temperature.

We comment that there is a soluble limit in which one can show
that only straight Polyakov loops control deconfinement.
For a theory on a sphere in the three spatial dimensions,
the radius $R$ of the sphere can be made of femtometer dimensions, so
that the non-Abelian
coupling constant runs to small values.  This theory is soluble,
and has a true phase transition at infinite $\Nc$
\cite{small_sphere}.  Near $\Tc$, the deconfining transition
is controlled by the simplest Polyakov loop, a straight lop in the fundamental
representation.  At zero coupling, deconfinement occurs 
at a temperature $\Tc$, which is equal to a pure number times $1/R$;
at this point, the mass term
for this Polyakov loop, alone, changes sign, jumping to a value of $1/2$.
Although the full behavior of this model has not yet been
computed, since the only dimensional scale in the problem is the radius
of the sphere, it is most likely the semi-QGP only persists up to temperatures
a few times $1/R$; {\it i.e.}, $\pTsemi/\Tc$ is a number of order one.

There is a heuristic explanation as to why lattice simulations
appear to find that $\pTsemi$ is just a few times $\Tc$.  If
$\pTsemi \gg \Tc$, then given how the coupling constant in QCD runs
with temperature \cite{coupling}, there would have to be a perturbative
mechanism for generating the eigenvalue repulsion necessary in the 
semi-QGP \cite{loopRPb}.  This would indicate a perturbative instability
for QCD in a thermal bath, which seems unnatural.  Instead, lattice
simulations find that the deviations from conformality, apart from
the usual perturbative corrections from the conformal anomaly, are
nonperturbative, due to corrections $\sim 1/T^2$ \cite{loopRPb,ren_lat,cheng,detar}.
If so, in a narrow regime in temperature
it is most natural that there are only a few operators which
contribute, and that these few only involve straight Polyakov loops.

Thus our conclusions about the universality of the bleaching of color
are special to a semi-QGP which exists only in a narrow region of temperature.
We also do not imply that all properties of fields in the semi-QGP are
independent of their mass: only the bleaching of color.  
This result may be of significance for experiment.  One of the real
puzzles of the experimental data from the Relativistic Heavy Ion
Collider is that the behavior of heavy quarks, such as charm, appears
to be rather similar to that of light quarks:
see Refs. \cite{whitepaper,strong}, and especially Ref. \cite{strong_rdp}.
This is very difficult to understand if the behavior arises from
energy loss, which is very different for light quarks than for heavy.
Our analysis suggests a completely different mechanism may be responsible:
the bleaching of color.  Of course a more careful analysis is necessary
in order to confirm this suggestion.

If RHIC is in a conformally
invariant regime, as suggested by
${\cal N}= 4$ supersymmetric gauge theories,
then results for heavy ions
at the LHC should be similar to RHIC: $\eta/s$ remains
small, essentially unchanged with temperature, and the behavior of
heavy quarks will remain like that of light quarks.

In contrast, if the semi-QGP is valid, 
then there will be dramatic differences between RHIC and the LHC.
If experiments at RHIC probe a region near $\Tc$, then those at the
LHC should probe temperatures significantly, perhaps a factor of two,
higher.  At the LHC, the ratio of the shear viscosity, to the entropy density,
will be large.  Also, in a 
perturbative QGP, the behavior of heavy quarks should differ significantly
from that of light quarks.  The difficulty is that these differences
are only for initial times and temperatures; inevitably, a system at
the LHC, even if it starts in the perturbative QGP, cools through the
semi-QGP.  

In the end, we eagerly await the experimental results for heavy ions
from the LHC, which will decide which theory is correct.

\acknowledgments
We would thank Olaf Kaczmarek and Kay H\"ubner for providing the lattice data.
This research of Y.H. was supported by the Grant-in-Aid for
the Global COE Program ``The Next Generation of Physics, 
Spun from Universality and Emergence'' from the Ministry of 
Education, Culture, Sports, Science and Technology (MEXT) of Japan.
This research of R.D.P. was supported 
by the U.S. Department of Energy under
cooperative research agreement \#DE-AC02-98CH10886.
R.D.P. also thanks
the Alexander von Humboldt Foundation for their support.

\end{document}